\documentclass[11pt]{article}
\usepackage{amsmath,amssymb}
\usepackage{amsfonts}
\usepackage{eucal}
\title{\leavevmode\vadjust{\vskip -5mm}
GEOMETRIC HAMILTON--JACOBI THEORY}
\author{\sc Jos\'e F. Cari\~nena\thanks{{\bf e}-{\it mail}: jfc@unizar.es}\\
 \tabaddress{Departamento de F\'{\i}sica Te\'orica. \\
Facultad de Ciencias, 
Universidad de Zaragoza, 
50009 Zaragoza. Spain}
\\
{\sc Xavier Gr\`acia\thanks{{\bf e}-{\it mail}: xgracia@ma4.upc.edu}}\\
 \tabaddress{Departament de Matem\`atica Aplicada IV, 
Universitat Polit\`ecnica de Catalunya\\
Campus Nord UPC edifici~C3, 
C. Jordi Girona~1, 08034 Barcelona, Catalonia, Spain}
\\
{\sc Giuseppe Marmo\thanks{{\bf e}-{\it mail}: giuseppe.marmo@na.infn.it}}\\
 \tabaddress{Dipartimento  di Scienze Fisiche, 
Universit\'a Federico II di Napoli, and INFN, Sezione di Napoli. \\
Complesso Univ.\ di Monte Sant'Angelo, Via Cintia, 
80126 Napoli. Italy}
\\
{\sc Eduardo Mart\'{\i}nez\thanks{{\bf e}-{\it mail}: emf@unizar.es}}\\
 \tabaddress{Departamento de Matem\'atica Aplicada. \\
Facultad de Ciencias, Universidad de Zaragoza, 50009 Zaragoza. Spain}
\\
{\sc Miguel C. Mu\~noz--Lecanda\thanks{{\bf e}-{\it mail}:
 matmcml@ma4.upc.edu}},
{\sc Narciso Rom\'an--Roy\thanks{{\bf e}-{\it mail}: nrr@ma4.upc.edu}}\\
 \tabaddress{Departamento de Matem\'atica Aplicada 4. \\
  Edificio C-3, Campus Norte UPC.
  C/ Jordi Girona 1. E-08034 Barcelona. Spain}}

\date{April 21, 2006}
\def\tabaddress#1{{\small\it\begin{tabular}[t]{c}#1
\\[1.2ex]\end{tabular}}}
\def\R{\mathbb{R}}

\def\Tr{\mathop{\rm Tr}\nolimits}
\def\pd#1#2{\frac{\partial #1}{\partial#2}}

\def\<#1>{\langle#1\rangle}
\def\gX{g(X)}    
\def\gX{X^\flat} 
\def\df{\Omega}
\def\Ker{\mathop{\rm Ker}\nolimits}
\def\derpar#1#2{\frac{\partial{#1}}{\partial{#2}}}
\newcommand{\vectorfields}[1]{\mathfrak{X}(#1)}

\newcommand{\cinfty}[1]{C^\infty(#1)}
\newcommand{\qquand}{\qquad\text{and}\qquad}
\newcommand{\FL}{\mathcal{F}L}
\newcommand{\FH}{\mathcal{F}H}

\newcommand{\g}{\mathfrak{g}}

\newtheorem{assumption}{Assumption}
\newtheorem{theorem}{Theorem}
\newtheorem{corollary}{Corollary}
\newtheorem{proposition}{Proposition}
\newtheorem{definition}{Definition}

\def\beq{\begin{equation}}
\def\eeq{\end{equation}}
\def\bea{\begin{eqnarray}}
\def\eea{\end{eqnarray}}
\def\beann{\begin{eqnarray*}}
\def\eeann{\end{eqnarray*}}
\def\ben{\begin{enumerate}}
\def\een{\end{enumerate}}
\def\proof{{\sl Proof}\quad}
\def\qed{\ifvmode\removelastskip\fi
{\unskip\nobreak\hfil\penalty50\hbox{}\nobreak\hfil \hbox{\vrule
height1.2ex width1.2ex}\parfillskip=0pt \finalhyphendemerits=0
\par\smallskip}}
\def\Tan{T}
\def\d{d}
\def\vl{\mathrm{vl}}
\let\ds=\displaystyle

\newcommand{\map}[3]{#1\colon#2\to#3}
\newcommand{\Real}{\mathbb{R}}

\def\texthook{\vrule height 0pt depth 0.4pt width 3.5pt
	      \vrule height 5pt depth 0.4pt \kern 3pt}
\def\scripthook{\vrule height 0pt depth 0.2pt width 1.5pt
                \vrule height 3pt depth 0.2pt width 0.2pt \kern 1pt}
\newcommand{\sode}{\textsc{sode}}
\newcommand{\cala}{\mathcal{A}}
\newcommand{\set}[2]{\left\{\,#1\left.\vphantom{#1#2}\,\right\vert\,#2\,\right\}} 
\newcommand{\quand}{\quad\mbox{and}\quad}
\newcommand{\pai}[2]{\langle#1,#2\rangle}
\newcommand{\calb}{\mathcal{B}}
\let\vf\vectorfields
\newcommand{\FhL}{\mathcal{F}_{\hat{L}}}
\newcommand{\hFL}{\hat{\mathcal{F}}_{L}}
\newcommand{\UX}{{\hat{X}}}
\newcommand{\gUX}{{\hat{X}^\flat}}
\newcommand{\at}[1]{\Big|_{#1}}

\begin{document}

\maketitle

\thispagestyle{empty}

\begin{abstract}
\noindent
The Hamilton--Jacobi problem is revisited bearing in mind the consequences 
arising from a possible bi-Hamiltonian structure. The problem is formulated 
on the tangent bundle for Lagrangian systems  in order to avoid 
the bias of the existence of a natural symplectic structure on the cotangent bundle. 
First it is developed for systems described by regular Lagrangians 
and then extended to systems described by singular Lagrangians with 
no secondary constraints. We also consider the example of the free relativistic particle,
the rigid body and the electron-monopole system.
\end{abstract}

\medskip
\noindent
{\sl Key words}: 
Hamilton--Jacobi equation, Lagrangian system, 
Hamiltonian system, singular dynamics, alternative Lagrangians.

\noindent
{\sl Mathematics Subject Classification (2000)}:
70H20, 70G45, 70H45, 70H03, 70H05
\\
{\sl PACS number (2003)}: 
02.40.Yy, 45.20.Jj

\clearpage
\tableofcontents
\clearpage

\section{Introduction}

Hamilton--Jacobi theory provides important physical examples of the
deep connection between first-order partial differential equations and
systems of first-order ordinary differential equations. In this respect,
it is also a stepping stone to the Schr\"odinger wave equation 
in quantum mechanics, and takes us as close as possible, 
within classical theory, to the notions of wave function and state in quantum theory.
 As a matter of fact, we obtain Hamilton--Jacobi-type equations whenever 
we consider a short-wave approximation for the solutions of wave-type equations, 
i.e., hyperbolic-type
equations (this includes the classical limit of quantum mechanics in the
Schr\"odinger picture by means of eikonal coordinates and the geometrical
optics limit of wave optics) \cite{EMS04}. Within the framework of wave mechanics, a
complete solution of the Hamilton--Jacobi equation allows us 
to reconstruct an approximate solution of the Schr\"odinger equation 
by providing us with the
phase of the wave function and the amplitude via the van Vleck determinant
constructed out of the Hessian of the complete solution itself 
(see, for instance, \cite{EMS04}, p. 172).

As regards the Hamiltonian formulation of geometrical optics, one may
recall that the origin of the whole method of canonical transformations in
analytical mechanics can be traced back to the famous memoirs on Optics
presented by  Hamilton to the Royal Irish Academy. There,
Hamilton showed
that the propagation of wavefronts can be entirely characterized by the
knowledge of a single function called the characteristic function. He also
showed that the characteristic function obeys a first order partial
differential equation, the so-called eikonal equation which is strictly
related to the Hamilton--Jacobi equation.

The name Hamilton--Jacobi is justified by the contribution given by Jacobi
that the dynamical problem (the ordinary differential equation) is completely
solved once a complete solution of the associated partial differential
equation is known.

Taking into account the fact that these equations were discovered almost two
centuries ago, one may believe that everything must be known for them. As a
matter of fact, we will argue and show in this paper, and in the forthcoming
ones, that there are several aspects which have so far not been considered.
Our own interest in reconsidering the Hamilton--Jacobi theory was
generated by the existence of bi-Hamiltonian descriptions for completely
integrable dynamical systems and the desire to unveil and understand the
quantum counterpart of bi-Hamiltonian systems. In particular, due to the
relevant role of the Hamilton--Jacobi theory in the Schr\"odinger
picture, it seems appropriate to achieve a proper understanding of the
Hamilton--Jacobi formulation for bi-Hamiltonian systems as a preliminary step
toward the the possibility of a better understanding of 
the corresponding quantum situation.

Vinogradov \cite{VK-75} has exhibited a deep relation between the commutation relations 
of differential operators acting on functions over the configuration space $Q$
and the canonical Poisson brackets of their principal symbols 
on the cotangent bundle $T^*Q$.
This connection seems to rule out the possibility of considering 
the Hamilton--Jacobi version of the bi-Hamiltonian systems. To escape this
apparent impossibility we find convenient to formulate 
the Hamilton--Jacobi theory on
the tangent bundle $TQ$  with the help of a regular
Lagrangian function and the associated Lagrangian two-form.
Thus we remove the bias of a  natural
symplectic structure on our carrier space, unlike in the
case of the cotangent bundle. Working with Lagrangians on $TQ$ we have the
possibility of dealing more directly with relativistic aspects and
with dynamical systems described by degenerate
Lagrangians (gauge theories), and therefore with the classical limit of their
corresponding quantum systems.

In this paper we will not address the problem
of constrained Lagrangian dynamics in full generality; this will be done in a
forthcoming paper. Subsequently we also shall consider the classical limit of
quantum bi-Hamiltonian systems and extend our Hamilton--Jacobi picture to
classical field theories which allow for bi-Hamiltonian descriptions.

As a spin-off from our tangent bundle formulation of the Hamilton--Jacobi 
problem, we will identify two main geometric aspects of the classical
formulation: the first one consists of finding a foliation transverse to the
 fibers  (of $TQ$ or $T^*Q$) and invariant under the dynamical evolution, while
the second one requires that the foliation be Lagrangian
with respect to a dynamically preserved symplectic structure. In this approach
the dynamics (ordinary differential equation) plays a prominent role
because we consider alternative Lagrangian or Hamiltonian
descriptions. Therefore, our generalization is to search for 
invariant foliations of the carrier
space with leaves having the same dimension as the configuration space $Q$,
since we drop the requirement of ``Lagrangianity''. 
Thus the partial differential equation associated with our problem
(equation (\ref{XGammaL}) in Section~\ref{LRHJ}) will be a
partial differential equation for a vector valued function rather than for a
scalar valued function as in the standard formulation. The transition
 from the vector valued function to the scalar valued  one takes place
with the help of the symplectic structure which allows us
to associate a closed 1-form (and therefore locally a function)
with our vector field by requiring the foliation to be Lagrangian.

When considering geodetical motions on Lie groups,
an interesting situation arises in which the first step is accomplished but the second
one is problematic, as we will see in Section 7.
Similar aspects also emerge when dealing with
bi-Hamitonian systems or systems described by equivalent Lagrangians: Here we
find invariant foliations; they may be Lagrangian with respect to one
symplectic structure but not Lagrangian with respect to some other
invariant symplectic structure.  We shall discuss a few very simple examples
to illustrate what is taking place.

The paper is organized as follows.
In Sections \ref{LRHJ} and \ref{Hamsec} we state the
Lagrangian and Hamiltonian geometrical formulations of the Hamilton--Jacobi problem,
respectively, showing how the standard classical problem
is a particular case of the extended one,
and clarifying the geometrical meaning of particular
and complete solutions~\cite{MMM90, Samuel}. 
The relation between both formalisms is also discussed.
Section \ref{uncons} is devoted to extending the theory
to the particular case of singular dynamical systems:
those where there are no Lagrangian constraints or,
what is equivalent, when secondary Hamiltonian constraints do not appear.
As an application of the above case,
the Hamilton--Jacobi problem for non-autonomous
Lagrangian and Hamiltonian systems is discussed in Section \ref{td}.
Finally, as examples, we apply our theory to the
free relativistic particle in Section \ref{frp},
to the free motion on a Lie group, to the rigid body, and to 
the electron-monopole system in Section~\ref{ems}.

\paragraph{Notation:}
Throughout this paper $Q$ is a $n$-dimensional differentiable manifold
representing the configuration space of a dynamical system,
and $\tau_Q\colon TQ\to Q$ and $\pi_Q\colon T^*Q\to Q$
are its tangent and cotangent bundles,
representing the phase spaces of velocities and momenta, respectively.

On the cotangent bundle there is a canonical symplectic form $\omega=-d\theta$,
where $\theta$ is the 1-form $\theta$ with coordinate expression $\theta=p_idq^i$.
Here and in the rest of this paper, sum over paired covariant and contravariant 
indices is understood. This symplectic form associates a vector field $Z_H$
to every function $H\in\cinfty{T^*Q}$, as the solution of the equation
$i(Z_H)\omega=dH$ (see e.g.~\cite{Ab} for the details). 

On the tangent bundle, the canonical object is the vertical endomorphism $S$,
with coordinate expression $S=(\partial/\partial v^i)\otimes dq^i$.
Given a Lagrangian function $L\in\cinfty{TQ}$,
we define the Cartan 1-form $\theta_L=d_SL=(\partial L/\partial v^i)dq^i$
and the Cartan 2-form $\omega_L=-d\theta_L$.
The dynamical vector fields associated to the Lagrangian
are the solutions of the dynamical equation $i(\Gamma)\omega_L=dE_L$,
where $E_L=\Delta(L)-L\in\cinfty{TQ}$ is the Lagrangian energy function
and $\Delta\in\vectorfields{TQ}$ is the Liouville vector field
(see e.g. \cite{Klein63,Cr81} and references therein).

We also remark that $T(TQ)$ has two different vector bundle structures
over $TQ$, given respectively by $\tau_{TQ} \colon T(TQ)\to TQ$, i.e. considering 
$TQ$ as new configuration space, and $T\tau_{Q} \colon T(TQ)\to TQ$. 
Maps $X \colon Q\to TQ$ that are sections for $\tau_Q$ are the vector fields in $Q$, 
and the set of such vector fields will be denoted by ${\mathfrak{X}}(Q)$.
Correspondingly, maps $X \colon TQ\to T(TQ)$ that are sections for $\tau_{TQ}$
are the vector fields in $TQ$, and those which
are also sections for $T\tau_{Q}$ are said to be second order differential equation fields
(hereafter referred to as {\sc sode} vector fields).
This means that their integral curves, which are the trajectories of the system, 
are holonomic. A vector field $X\in {\mathfrak{X}}(Q)$ can be lifted to $TQ$
producing the so called complete or tangent lift of $X$ and denoted by
$X^T \in {\mathfrak{X}}(TQ)$. More details can be found in \cite{Cr83,MaS85}.

\section{Lagrangian formulation of the Hamilton--Jacobi problem}
\protect\label{LRHJ}

In this section we formulate the Hamilton--Jacobi problem on the tangent bundle.
In this setting we are able to handle dynamical systems which admit
alternative Lagrangian descriptions, and we clearly show how the search
for solutions of the Hamilton--Jacobi problem splits in two steps.

We recall that, in the standard formulation, the Hamilton--Jacobi problem
consists in finding a function $S(t,q)$, known as the principal function,
such that the partial differential equation ({\sc pde})
\[
\pd S t+H\left(q,\pd{S}{q}\right)=0\,,
\]
 is satisfied. If we put $S(t,q)=W(q)-t\,E$, where $E$ is a constant,
then the function $W$, known as the characteristic function, has to satisfy
\begin{equation}
\label{HJ-local}
H\left(q,\pd{W}{q}\right)=E\,.
\end{equation}
Both of the above \textsc{pde} are known as the Hamilton--Jacobi equation.
However we will always refer to the second one.

In geometric terms, equation~\eqref{HJ-local} can be written as $(dW)^*H=E$,
where we understand $dW$ as a section of the cotangent bundle.
In other words, we look for a section $\alpha$ of $T^*Q$ such that $\alpha^*H=E$
and $\alpha$ is a closed 1-form, $d\alpha=0$, and hence locally exact, $\alpha=dW$.
The second condition, $d\alpha=0$, can alternatively be expressed in terms of
the canonical symplectic form on $T^*Q$ in the form $\alpha^*\omega=0$,
so that one can reformulate the Hamilton--Jacobi equation in the form~\cite{Ab} 
\begin{equation}
\label{HJ-H}
\alpha^*H=E,\qquad \alpha^*\omega=0\,.
\end{equation}

Consider now the Lagrangian formalism. Let $L\in\cinfty{TQ}$
be the Lagrangian function and $\theta_L$, $\omega_L$ be the associated Cartan forms.
A literal translation of the above coordinate-free formulation of
the Hamilton--Jacobi equations from the cotangent
bundle to the tangent bundle would be~\cite{Prince}
\begin{equation}
\label{HJ-L}
X^*(E_L)=E\,,\qquad X^*(\omega_L)=0\,,
\end{equation}
where $X:Q\to TQ$ is the unknown ``vector valued'' function, and the second
equation states that the vector field $X$ is associated (at least locally)
with a function $W$ by means of the relation $X^*(\theta_L)=dW$,
which is a stronger version of $X^*(\omega_L)=0$.

Among the many important consequences that may be deduced from
the existence of a solution of the Hamilton--Jacobi equation,
let us recall the following. Let $P(q)=\pd{W}{q}(q)$,
and consider the vector field $X=\pd{H}{q}(q,P(q))$.
If $q=\gamma(t)$
is a solution of the differential equation $\dot{q}=X(q)$,
then $\lambda(t)=(\gamma(t),P(\gamma(t)))$ is a solution of the Hamilton equations.
The Lagrangian counterpart of this property reads as follows.
If $X$ is a solution of~\eqref{HJ-L} and $q=\gamma(t)$ is a solution
of the differential equation $\dot{q}=X(q)$, then $\xi(t)=(\gamma(t),X(\gamma(t)))$
is a solution of the Euler--Lagrange equations.

This fact will be our starting point in the study of the Hamilton--Jacobi equation
and its generalization. We will look for the implications of this property
and its relation with equations~\eqref{HJ-L}.

\subsection{Statement of the problem and solutions}
\label{statementoftheproblem}

We will assume first that the Lagrangian $L$ is regular,
and we will leave for Section~\ref{uncons} the analysis of the unconstrained singular case.
The regularity of the Lagrangian is equivalent to the regularity of the Cartan 2-form,
so that $\omega_L$ is symplectic. It follows that there exists
a unique solution $\Gamma_L\in\vectorfields{TQ}$ of the Lagrangian dynamical equation
\begin{equation}
i(\Gamma_L)\omega_L=dE_L \ .
\label{ELeq}
\end{equation}
$\Gamma_L$ is called the Lagrangian vector field of the Lagrangian system.
It is well known~\cite{Cr81} that $\Gamma_L$ is a second order differential equation.

\paragraph{Generalized Lagrangian Hamilton--Jacobi problem.}
{\it 
Let $L\in C^\infty(TQ)$ be a Lagrangian function. The 
{\rm generalized Lagrangian Hamilton--Jacobi problem} consists 
in finding a vector field $X\colon Q\to TQ$ 
such that, if $\gamma\colon\R\to Q$ is an integral curve
of $X$, i.e. $\dot\gamma=X\circ \gamma$,  then $\dot\gamma\colon\R\to TQ$
is an integral curve of $\Gamma_L$; that is,
$$
X\circ\gamma=\dot\gamma \ \Longrightarrow \
\Gamma_L\circ\dot\gamma=\dot{\overline{X\circ\gamma}} \ .
$$
$X$ is said to be a {\rm solution of the generalized 
Lagrangian Hamilton--Jacobi problem}.
}

\medskip

As we will see in a moment, in geometrical terms, this requirement means that
the image of $X$, as a map from $Q$ to $TQ$, is a $\Gamma_L$-invariant submanifold of $TQ$.
Let us show first an example.

\paragraph{Example 1}

The dynamics of the free particle in $\R^2$ is given
by the regular Lagrangian function 
$$
L(q^1,q^2,v^1,v^2)=\frac 12\left[(v^1)^2+(v^2)^2\right]\ ,
$$
with associated geometrical objects
\begin{eqnarray}
\theta_L&=&\pd L{v^1}\,dq^1+\pd L{v^2}\,dq^2=v^1\, dq^1+v^2\, dq^2\cr
E_L&=&\frac 12\left[(v^1)^2+(v^2)^2\right]\cr
\omega_L&=&dq^1\wedge dv^1+dq^2\wedge dv^2\cr
\Gamma_L&=&v^1\,\pd{}{q^1}+v^2\,\pd{}{q^2} \nonumber
\end{eqnarray}
The vector field
$$
X=k\,\pd{}{q^1}+\frac {k\,q^2-l}{q^1}\, \pd {}{q^2}\,,\qquad k,l\in \R\,,
$$
defines a two-parameter family of vector fields on $Q=\R^2$ which are
generalized solutions. We also find that 
\begin{eqnarray}
X^*(\omega_L)&=&-\frac{k\,q^2-l}{q^1}\,dq^1\wedge dq^2\,,\cr
X^*(E_L)&=&\frac 12 \left[k^2+\left(\frac{k\,q^2-l}{q^2}
\right)^2
\right]\,.\nonumber
\end{eqnarray}

Thus, the simple translation of the geometrical relations from $T^*Q$ to $TQ$
would be violated. Now we can formulate on $TQ$ a {\sc pde} which replaces the {\sc pde}
for the characteristic function $W$. We find that it must be stated in terms
of a vector valued function.

\begin{proposition}
$X$ is a solution of the generalized Lagrangian
Hamilton--Jacobi problem if, and only if,
$X$ and $\Gamma_L$ are $X$-related; that is,
\begin{equation}
\Gamma_L\circ X=TX\circ X \ .
\label{XGammaL}
\end{equation}
\label{XGammaLprop}
\end{proposition}
\proof
$X$ is a solution of the generalized Lagrangian Hamilton--Jacobi problem
if, for every $\gamma\colon\R\to Q$ such that
$X\circ\gamma=\dot\gamma$, then 
$$
\Gamma_L\circ\dot\gamma=\dot{\overline{X\circ\gamma}}=
TX\circ\dot\gamma=TX\circ X\circ\gamma \ .
$$
But $\Gamma_L\circ\dot\gamma=\Gamma_L\circ X\circ\gamma$, 
and as $X$ has integral curves through every point
$q\in Q$, this is equivalent to
$TX\circ X=\Gamma_L\circ X$.

The proof of the converse is straightforward.
\qed

This equation for a given 
{\sc sode} $\Gamma_L$ defines a {\sc pde} for $X$ and replaces the {\sc pde} for $W$ in the
standard formulation of the Hamilton--Jacobi problem.

In addition we have:

\begin{proposition}
$X$ is a solution of the generalized
Lagrangian Hamilton--Jacobi problem
if, and only if, the submanifold ${\rm Im}\,X \subset TQ$
is invariant by the Lagrangian vector field $\Gamma_L$
(that is, $\Gamma_L$ is tangent to the submanifold $X(Q)$).
\label{Xinvar}
\end{proposition}
\proof
For the direct implication, 
it suffices to show that, for every $q\in Q$,
$\Gamma_L(X_q)$ is tangent to ${\rm Im}\, X$,
and it holds because,
by proposition \ref{XGammaLprop}, 
$\Gamma_L(X_q)=T_qX(X_q)$.

Conversely, if $\Gamma_L$ leaves ${\rm Im}\, X$ invariant, then
$\Gamma_L(X_q)\in T_{X_q}{\rm Im}\, X$.
Therefore, there exists $u \in T_qQ$
such that $\Gamma_L(X_q)=T_qX(u)$; hence
$$
X_q=(T\tau_Q\circ\Gamma_L)(X_q)=(T_q\tau_Q\circ T_qX)(u)=T_q(\tau_Q\circ X)(u)=u\ ,
$$
because $\tau_Q\circ X={\rm Id}_Q$, and $\Gamma_L$ being a \textsc{sode}, it is a section of the projection $T\tau_Q$, so $T\tau_Q\circ\Gamma_L={\rm Id}_{TQ}$.
Thus $\Gamma_L(X_q)=T_qX(X_q)$ for every $q\in Q$; that is, $\Gamma_L\circ X=TX\circ X$,
and $X$ is a solution of the generalized
Lagrangian Hamilton--Jacobi problem by proposition \ref{XGammaLprop}.
\qed

If $X$ is a solution of the generalized
Lagrangian Hamilton--Jacobi problem, then
the integral curves of $X$ are the $\tau_Q$-projection
of integral curves of $\Gamma_L$ contained in ${\rm Im}\, X$.

Observe that we have not used that $\Gamma_L$ is the Lagrangian vector field,
so these results actually hold  for every \textsc{sode} $\Gamma\in\vectorfields{TQ}$.
Using the fact that $\Gamma_L$ is the Lagrangian
vector field of a Lagrangian system,  the
above results can be related with the energy Lagrangian function $E_L$
in the following way, which avoids the explicit calculation of the
dynamical Lagrangian vector field.

\begin{theorem}
$X$ is a solution of the generalized Lagrangian
Hamilton--Jacobi problem if, and only if,
\begin{equation}
i(X)(X^*\omega_L)=d(X^*E_L)
\label{1}
\end{equation}
\label{proplag}
\end{theorem}
\proof
 From the Lagrangian dynamical equation (\ref{ELeq}) we obtain
$$
X^*i(\Gamma_L)\omega_L=X^*dE_L=d(X^*E_L) \ ,
$$
but, as $X$ and $\Gamma_L$ are $X$-related (proposition \ref{XGammaLprop}), 
we have that
$$
X^*i(\Gamma_L)\omega_L=i(X)(X^*\omega_L) \ ,
$$
which yields (\ref{1}).

Conversely, suppose that $X$ satisfies (\ref{1}). The deviation $D_L$ from the relatedness
$$
D_L=\Gamma_L\circ X-TX\circ X\colon Q\to TTQ \ ,
$$
is a vector field along $X$.
We have to prove that $D_L=0$.
First we have that $D_L$ is $\tau_Q$-vertical. In fact,
$\tau_Q\circ X={\rm Id}_Q$, and $\Gamma_L$ being a {\sc sode},
it is a section of the projection $T\tau_Q$, so
$T\tau_Q\circ\Gamma_L={\rm Id}_{TQ}$, hence
$$
T\tau_Q\circ D_L=
T\tau_Q\circ\Gamma_L\circ X-T\tau_Q\circ TX\circ X=X-X=0 \ .
$$
Furthermore, from the Lagrangian dynamical equation (\ref{ELeq}) 
we have $X^*i(\Gamma_L)\omega_L=X^*dE_L=d(X^*E_L)$, which
combined with the hypothesis, $i(X)(X^*\omega_L)=d(X^*E_L)$,
leads to $X^*i(\Gamma_L)\omega_L-i(X)(X^*\omega_L)=0$.
Therefore, for every $q\in Q$ and $Y_q\in T_qQ$, we have
\beann
0 &=& (X^*i(\Gamma_L)\omega_L-i(X)(X^*\omega_L))_q(Y_q)=
(\omega_L)_{X(q)}(\Gamma_L(q),T_qX(Y_q))-(X^*\omega_L)_q(X_q,Y_q)
\\ &=&
(\omega_L)_{X(q)}(\Gamma_L(q),T_qX(Y_q))-
(\omega_L)_{X(q)}(T_qX(X_q),T_qX(Y_q))
\\ &=&
(\omega_L)_{X(q)}(D_L(q),T_qX(Y_q)) \ .
\eeann
Moreover, for every $\tau_Q$-vertical
vector field $V\in\vectorfields{TQ}$ we have that
$$
(\omega_L)_{\alpha(q)}(D_L(q),V(X(q)))=-d(\theta_L)_{\alpha(q)}
(D_L(q),V(X(q))) \ ,
$$
which vanishes for every $q\in Q$.
 We recall that if $\alpha$ is a semibasic form, then $d\alpha(V_1,V_2)=0$
for every pair of vertical fields $V_1$ and $V_2$. But 
 $\theta_L$ is a $\tau_Q$-semibasic form, 
and  $T_{X(q)}TQ=T_{X(q)}({\rm Im}\, X)\oplus{\rm V}_{X(q)}(\tau_Q)$,
thus we have proved that
$$
(\omega_L)_{\alpha(q)}(D_L(q),Z(X(q)))=0
\ , \ 
\mbox{\rm for every $q\in Q$, $Z\in\vectorfields{TQ}$} \ ,
$$
and hence $D_L(q)=0$, for every $q\in Q$, since $\omega_L$ is nondegenerate.
\qed

To solve the generalized Lagrangian Hamilton--Jacobi problem is,
in general, a hard task; it amounts to finding $\Gamma_L$-invariant submanifolds of $TQ$
which are transverse to the fibers. Thus, it is convenient to consider
a less general problem, which constitutes the standard version of the Lagrangian
Hamilton--Jacobi problem:

\paragraph{Lagrangian Hamilton--Jacobi problem}
{\it
Given a Lagrangian function $L\in C^\infty(TQ)$, the 
{\rm Lagrangian Hamilton--Jacobi problem} consists 
in finding solutions $X$ of the generalized 
Lagrangian Hamilton--Jacobi problem satisfying that $X^*\omega_L=0$.
}

\medskip

As $0=X^*\omega_L=-X^*d\theta_L=-d(X^*\theta_L)$,
we have that every point has an open neighborhood $U\subset Q$
where there is a function $W\in\cinfty{U}$ such that $X^*\theta_L=dW$ (in $U$).

{\bf Remark:} 
In the example of the free particle in $\R^2$ given above,
the pull-back of the symplectic 2-form $\omega_L$ by the vector field
$X$ is different from zero.  Hence, $X$ does not provide a solution
of the Lagrangian Hamilton--Jacobi problem because it is not associated with
a closed 1-form on the configuration space.

\medskip

A straightforward consequence of the last theorem is:

\begin{corollary}
If $X$ is a solution of the Lagrangian
Hamilton--Jacobi problem, then $d(X^*E_L)=0$.
\label{coro2}
\qed
\end{corollary}

Observe that 
if $X$ is a solution of the Lagrangian
Hamilton--Jacobi problem, then ${\rm Im}\,X$ 
is a Lagrangian submanifold
of $(TQ,\omega_L)$ contained in a level set of~$E_L$. 
In fact, $\dim\,{\rm Im}\,X=n$ and,
if $j_X\colon{\rm Im}\,X\hookrightarrow TQ$
denotes the natural embedding, 
we have that $j_X^*\omega_L=0$,
due to $X^*(\omega_L)=0$.

We can summarize the above results in the following:

\begin{proposition}
Let $X\in\vectorfields{Q}$ satisfy
$X^*\omega_L=0$. 
Then, the following assertions are equivalent:
\begin{enumerate}
\item 
$X$ is a solution of the Lagrangian Hamilton--Jacobi problem.
\item
$d(X^*E_L)=0$
\item
${\rm Im}\, X$ is a Lagrangian submanifold of $TQ$ invariant 
by~$\Gamma_L$.
\item
The integral curves of $\Gamma_L$ with initial conditions
in ${\rm Im}\, X$ project onto the integral curves of~$X$.
\end{enumerate} 
\qed
\label{Lagconverse}
\end{proposition}

\paragraph{Coordinate expressions}

Let us show the local expressions of the objects so far presented.
Consider coordinates $(q^i)$ on~$Q$, 
and the corresponding natural coordinates 
$(q^i,v^i)$ on its tangent bundle.

Consider an arbitrary vector field $\Gamma \in \vectorfields{\Tan Q}$
satisfying the second-order condition,
$\Gamma(q,v)=(q,v;v,a(q,v))$,
and a vector field $X \in \vectorfields{Q}$: $X(q)=(q,w(q))$.
Then we have
\beq
(\Tan X \circ X - \Gamma \circ X)(q) =
\left(
q, w(q); 0, \frac{\partial w}{\partial q} w - a(q,w(q))
\right) ,
\eeq
which is a vertical vector field along~$X$.
Its vanishing is the necessary and sufficient
condition for $X$ and $\Gamma$ to be $X$-related:
$$
\frac{\partial w^i}{\partial q^j} w^j(q) - a^i(q,w(q)) = 0.
$$
This equation is the \textsc{pde} for the vector valued function $w^j(q)$ which replaces
the standard \textsc{pde} for the scalar function $W$.

When $\Gamma$ is the Lagrangian vector field~$\Gamma_L$,
its components satisfy 
$\ds
W_{ij} a^j = 
\frac{\partial L}{\partial q^i} - 
\frac{\partial^2 L}{\partial v^i\,\partial q^j}v^j
$,
where 
$\ds
W_{ij}= \frac{\partial^2 L}{\partial v^i\partial v^j}
$
is the Hessian matrix of~$L$.

Then we can compute the 1-form
\begin{equation}
\left.
-i_X X^*(\omega_L) + X^*(\d E_L) =
\left(
\frac{\partial^2 L}{\partial v^i\partial v^j}
\frac{\partial w^j}{\partial q^k} w^k +
\frac{\partial^2 L}{\partial v^i\partial q^j} w^j -
\frac{\partial L}{\partial q^i}
\right)\right|_{v=w(q)} 
\d q^i ,
\end{equation}
whose vanishing also expresses that $X$ is a 
solution of the generalized Lagrangian Hamilton--Jacobi problem.

Looking carefully at the local expressions one can find a relation
between 
$-i_X X^*(\omega_L) + X^*(\d E_L)$
and
$\Tan X \circ X - \Gamma \circ X$, which is given by the Hessian, as we are going to show.

To this end, let us first recall that, for any vector bundle $E \to Q$,
we have the vertical lift map $\vl_E \colon E \times_Q E \to \mathrm{V} E \subset \Tan E$, 
an isomorphism which in fiber coordinates reads 
$\vl(q,u,v) = (q,u;0,v)$. With $E=TQ$, this gives an isomorphism
$\vl\colon{TQ\times_Q TQ} \to {V(TQ)\subset T(TQ)}$.

Associated with the Lagrangian $L$, we have the Legendre transformation
$\map{\FL}{TQ}{T^*Q}$, which in coordinates reads $\ds\FL(q,v)=(q^i,\pd{L}{v^i})$.
In other words, $\FL$ is the fibre derivative of $L$.

Moreover, we can define the fiber Hessian 
$\mathcal{F}^2L \colon \Tan Q \to \Tan^*Q \otimes \Tan^*Q$ which
defines, if the Lagrangian is regular,
another isomorphism
$
\widehat{\mathcal{F}^2L} \colon 
\Tan Q \times_Q \Tan Q \to \Tan Q \times_Q \Tan^*Q 
$.
In coordinates,
$\mathcal{F}^2L(q,v)=(q^i,W_{ij})$
and
$\widehat{\mathcal{F}^2L} (q^i,v^i,u^i) = (q^i,v^i,W_{ij}u^j)$.

With these ingredients, we achieve an alternative understanding
of Theorem~\ref{proplag}:
\begin{proposition}
Let $\vl$ be the vertical lift map
of the tangent bundle $\Tan Q$.
For any vector field $X$ on~$Q$,
we have
\beq
(X,-i_X X^*(\omega_L) + X^*(\d E_L))
=
\widehat{\mathcal{F}^2L} \circ \vl^{-1} \circ
(\Tan X \circ X - \Gamma_L \circ X) .
\eeq
Therefore, 
$-i_X X^*(\omega_L) + X^*(\d E_L)$ vanishes if and only if
$\Tan X \circ X - \Gamma_L \circ X$ vanishes too.
\qed
\end{proposition}

\subsection{Complete solutions}
\protect\label{cs}

The most useful and essential idea in the standard Hamilton--Jacobi theory
consists in finding, not only one particular solution as we have used
in the previous subsection, but rather a complete solution of the problem.
This may be defined as follows.

\begin{definition}
Consider a solution $X_\lambda$ depending on $n$ additional parameters
$\lambda\in\Lambda$, where $\Lambda\subseteq\R^n$ is an open set,
and suppose that the map $\Phi\colon Q\times\Lambda\to TQ$ given by
$\Phi(q,\lambda)=X_\lambda(q)$ is a local diffeomorphism. 
In this case $\{ X_\lambda; \lambda\in\Lambda\}$ 
is said to be a {\rm complete solution} of the generalized
Lagrangian Hamilton--Jacobi problem.
\end{definition}

 From the definition, it follows that a complete solution provides $TQ$
with a foliation transverse to the fibers, and that the Lagrangian vector field
$\Gamma_L$ is tangent to the leaves. 

If $\{ X_\lambda; \lambda\in\Lambda\}$ is a complete solution,
the integral curves of $X_\lambda$, for  different $\lambda\in\Lambda$,
will provide all the integral curves of the Lagrangian vector field $\Gamma_L$.
This means that, if $(q_0,v_0)\in{\rm Im}\, X$, then there is $\lambda_0\in\Lambda$
such that $X_{\lambda_0}(q_0)=v_0$, and the integral curve of $X_{\lambda_0}$
through $q_0$, lifted by $X_{\lambda_0}$ to $TQ$,
gives the integral curve of $\Gamma_L$ through $(q_0,v_0)$.
This justifies the name of ``complete solution''.

{\bf Remark:}  We may use instead a fiber bundle $P$ over $\Lambda$, such that
$\Gamma_L$ projects onto the null vector field; i.e. $\Lambda$ is a space of
constants of the motion and fibers have the same dimension as the configuration
space $Q$. Thus we may take into account the nontriviallity of $P$ as a
bundle. On the other hand, if $\Lambda$ were contractible  the bundle would be
trivial, and we would revert to the previous situation.

Furthermore, different transversal foliations of $TQ$,
with $\Gamma_L$ tangent to the leaves, are different ways to collect
solutions of $\Gamma_L$ smoothly and such that
they project onto $Q$ in a coherent way:
integral curves of $\Gamma_L$ in ${\rm Im}\,X_\lambda$
project onto integral curves of the associated vector field $X_\lambda$.

The relation between $\Gamma_L$ and complete solutions is the following:
\begin{itemize}
\item
If we have a family of $n$ first integrals $f_1,\ldots,f_n$
of $\Gamma_L$ such that $d_Sf_1\wedge\ldots\wedge d_Sf_n\not=0$,
then $f_i=c_i$, $c_i\in\Real$, for $i=1,\ldots, n$,
define a transversal foliation. Thus we can locally isolate the velocities
as functions of the coordinates $q^i$ and the constants $c_i$.
Now, replacing in the expression of $\Gamma_L$ these velocities
and projecting to the basis, we obtain a local complete solution
$X_{(c_1,\ldots,c_n)}$.
\item
Conversely, if $\Phi\colon Q\times\Lambda\to TQ$ is a complete solution,
then the functions defining locally the foliation give us the above family of integrals
of motion of $\Gamma_L$. More explicitly, the components of the map
$\map{F}{TQ}{\Lambda}$ given by $F=\mathrm{pr}_2\circ\Phi^{-1}$,
are constants of the motion.
\end{itemize}

Moreover, if the foliation is Lagrangian in $(TQ,\omega_L)$,
then we have a complete solution of the Lagrangian Hamilton--Jacobi problem.
In this case the above family of first integrals
are in involution.

In our previous example of the free particle, varying the parameters $(k,l)\in\R^2$
we obtain a complete solution.

All these considerations are shown in the following example.

\paragraph{Example 2}

Let us consider the example of the two-dimensional 
standard harmonic oscillator described by
$$
L=\frac{1}{2}((v^1)^2+(v^2)^2-(q^1)^2-(q^2)^2)\ .
$$
The dynamical vector field is 
$$
\Gamma_L=v^1\, \pd{}{q^1}+v^2\, \pd{}{q^2}-q^1\,\pd{}{v^1}-q^2\, \pd{}{v^2}\ ,
$$
and the standard Lagrangian symplectic 2-form is
$\omega_L=dq^1\wedge dv^1+dq^2\wedge dv^2$.

We know that the functions 
$$
f_1 = v^1 v^2 + q^1 q^2, \ 
f_2 = (v^1)^2 + (q^1)^2, \
f_3 = (v^2)^2 + (q^2)^2, \ 
f_4 = q^1 v^2 - q^2 v^1 
$$
are constants of the motion.
Of course, not all of them are functionally independent.
Suppose their values are 
$f_1=C$, $f_2=2E_1$,
$f_3=2E_2$, $f_4=l$.
We can use, for instance, $f_2$ and $f_3$ to express  $v^1$
and $v^2$ as functions of the base coordinates  and the two parameters 
$E_1$ and $E_2$, and using these expressions in the 
dynamical vector field we find a vector field on the base
$Q$ depending on the two energies:
$$
X_{E_1,E_2}=  \left(\pm\sqrt{2\,E_1-(q^1)^2}\, \pd {}{q^1} \pm
\sqrt{2\,E_2-(q^2)^2}\, \pd {}{q^2}\right)\ .
$$
Note that the two functions we have used are in involution,
$\{f_2,f_3\}=0$,
and that 
$$
(X_{E_1,E_2})^*\omega_L=dq^1\wedge d(\sqrt{2\,E_1-(q^1)^2})+
dq^2\wedge d(\sqrt{2\,E_2-(q^2)^2})=0\ .
$$
This is a 2-parameter family of vector fields, 
for which the images are Lagrangian submanifolds 
with respect to~$\omega_L$.

We can also choose the functions $f_1$ and $f_4$ for obtaining 
expressions of the velocities in terms of positions,
when $q^1v^2+q^2v^1\ne 0$, because
$$
d_Sf_1\wedge d_Sf_4=(v^2\, dq^1+v^1\, dq^2)\wedge (-q^2\, dq^1+q^1\, dq^2)
=(q^1v^2+q^2v^1)\, dq^1\wedge dq^2\ ,$$
and in this case,
$$
v^1=\frac{-l\pm \sqrt{l^2+4\,q^1\,q^2(C-q^1\,q^2)}}{2\, q^2}\,,\qquad
 v^2=\frac{l\pm \sqrt{l^2+4\,q^1\,q^2(C-q^1\,q^2)}}{2\, q^1}\,=
\frac{l+q^2v^1}{q^1}\ ,
$$
and we have the vector field in $Q$
\beann
X_{C,l}(q^1,q^2)&=&
\left(\frac{-l\pm \sqrt{l^2+4\,q^1\,q^2(C-q^1\,q^2)}}{2\, q^2}\right)
\,\pd{}{q^1}
\\  &+&
 \left(\frac{l\pm \sqrt{l^2+4\,q^1
\,q^2(C-q^1\,q^2)}}{2\, q^2}\right)\,\pd{}{q^2}\ .
\eeann
However, notice that because of 
$\{f_1,f_4\}=f_2-f_3$,
we find
$(X_{C,l})^*\omega_L\not =0$. Therefore, $X_{C,l}$
is a complete solution for the generalized problem, but not for the
standard Hamilton--Jacobi problem.

\section{Formulation of the Hamilton--Jacobi problem on $T^*Q$}
\protect\label{Hamsec}

\subsection{Statement of the problem and solutions}

We now consider the Hamiltonian formalism in the cotangent bundle.
Let $H\in\cinfty{T^*Q}$ be a Hamiltonian function, and denote by 
$\omega=-d\theta\in\df^2(T^*Q)$ the canonical symplectic form.
There exists a unique vector field $Z_H\in\vectorfields{T^*Q}$
whose integral curves are the trajectories of the system; that is,
the solutions of the Hamilton equation. Geometrically this means that
$Z_H$ is the solution of the Hamiltonian dynamical equation
\begin{equation}
i(Z_H)\omega=dH \ .
\label{Heq}
\end{equation}
$Z_H$ is called the Hamiltonian vector field of the system.

As in the Lagrangian formalism, let us start with
the generalized version of the  Hamilton--Jacobi problem,
which can be stated as follows:

\paragraph{Generalized Hamiltonian Hamilton--Jacobi problem}
{\it
Given a Hamiltonian vector field $Z_H\in \vectorfields{T^*Q}$, the 
{\rm generalized Hamiltonian Hamilton--Jacobi problem} consists in finding
a vector field $X\colon Q\to TQ$ and
a $1$-form $\alpha\colon Q\to T^*Q$
such that, if $\gamma\colon\R\to Q$ is an integral curve
of $X$, then $\alpha\circ\gamma\colon\R\to T^*Q$
is an integral curve of $Z_H$. That is,
\beq
X\circ\gamma=\dot\gamma \ \Longrightarrow \
\dot{\overline{\alpha\circ\gamma}}=Z_H\circ(\alpha\circ\gamma) \ .
\label{cond1}
\eeq
}

\medskip

The first result is:

\begin{proposition}
Given a vector field $Z_H\in \vectorfields{T^*Q}$, 
$(X,\alpha)$ satisfies the condition (\ref{cond1}) if, and only if,
the vector fields 
$X$ and $Z_H$ are $\alpha$-related; that is,
\begin{equation}
Z_H\circ\alpha=T\alpha\circ X \ .
\label{XZH}
\end{equation}
\label{XZHprop}
\end{proposition}
\proof
If $(X,\alpha)$ satisfies the condition (\ref{cond1}) then, for every
$\gamma\colon\R\to Q$ such that $X\circ\gamma=\dot\gamma$,
we have
$$
Z_H\circ\alpha\circ\gamma=\dot{\overline{\alpha\circ\gamma}}=
T\alpha\circ\dot\gamma=T\alpha\circ X\circ\gamma \ ;
$$
but, as $X$ has integral curves through every point
$q\in Q$, this condition is equivalent to
$Z_H\circ\alpha=T\alpha\circ X$.

The proof of the converse is straightforward.
\qed

In fact, both elements $(X,\alpha)$ satisfying the condition (\ref{cond1}) 
are related, since, by composing both sides of the above equation (\ref{XZH}) with $T\pi_Q$,
and taking into account that
$\pi_Q\circ\alpha={\rm Id}_Q$,
we have the following immediate consequence:

\begin{corollary}
If $(X,\alpha)$ satisfies the condition (\ref{cond1}) then
$$
X=T\pi_Q\circ Z_H\circ\alpha .
$$
\label{XZHcor}
\end{corollary}

It is interesting to remark that we also have 
the following relation between $X$ and $\alpha$:
$$
X=\FH\circ\alpha \ ,
$$ 
where $\FH\colon T^*Q\to TQ$ is the fiber derivative of the
Hamiltonian function.

In terms of our previous geometrical formulation, this amounts to $X^*(\theta_L)=\alpha$
when $L$ is the Lagrangian function associated with $H$.

As $X$ is determined by $\alpha$, we introduce the following:

\begin{definition}
A {\rm solution} of the generalized Hamiltonian Hamilton--Jacobi problem
for $Z_H$ is a $1$-form $\alpha\in\df^1(Q)$
such that, if $\gamma\colon\R\to Q$ is an integral curve
of $X=T\pi_Q\circ Z_H\circ\alpha$, then $\alpha\circ\gamma\colon\R\to T^*Q$
is an integral curve of $Z_H$; that is,
$$
T\pi_Q\circ Z_H\circ\alpha\circ\gamma=\dot\gamma \ \Longrightarrow \
\dot{\overline{\alpha\circ\gamma}}=Z_H\circ(\alpha\circ\gamma) \ .
$$
Then $X=T\pi_Q\circ Z_H\circ\alpha$ is said to be the
{\rm vector field associated with $\alpha$}.
\end{definition}

\paragraph{Example 1} (continued)

Consider the Hamiltonian function for a free particle in $\R^2$
$$
H(q^1,q^2,p_1,p_2)=\frac 12\left( {p_1}^2+{p_2}^2 \right) .
$$
The $1$-form $\ds \alpha=\frac{1}{q^1}\, dq^2$
and its associated vector field
$\ds X=\frac{1}{q^1}k_1\pd{}{q^2}$ provide a solution for
 the generalized Hamiltonian Hamilton--Jacobi problem,
but we will see later that they do not give rise to any solution of the
standard Hamilton--Jacobi problem.

\medskip

\begin{proposition}
Given a vector field $Z_H\in \vectorfields{T^*Q}$, 
a 1-form $\alpha\in\df^1(Q)$ is a solution of the generalized 
Hamiltonian Hamilton--Jacobi problem if, and only if, 
the submanifold ${\rm Im}\,\alpha \subset T^*Q$
is invariant under the flow of the vector field $Z_H$
(that is, $Z_H$ is tangent to the submanifold ${\rm Im}\,\alpha$).
\label{otrapropo}
\end{proposition}
\proof
If $\alpha\in\df^1(Q)$ is a solution of the
Hamiltonian Hamilton--Jacobi problem, and
$X=T\pi_Q\circ Z_H\circ\alpha$, then
$Z_H\circ\alpha=T\alpha\circ X$, and thus
$Z_H(\alpha(q))=T\alpha(X(q))$, for every $q\in Q$.
Hence, $Z_H$ is tangent to ${\rm Im}\,\alpha$.

Conversely,
if ${\rm Im}\,\alpha$ is invariant by $Z_H$ then
$Z_H(\alpha(q))\in T_{\alpha(q)}{\rm Im}\,\alpha$,
which implies that there exists $u\in T_qQ$ such that
$Z_H(\alpha(q))=T_q\alpha(u)$. Defining
$X$ by $T_q\alpha(X_q)=Z_H(\alpha(q))$, then
$X$ is differentiable since $X=T\pi_Q\circ Z_H\circ\alpha$.
Hence $X$ is a vector field in $Q$ which satisfies 
$Z_H\circ\alpha=T\alpha\circ X$, and then $\alpha$ is a solution of the
Hamiltonian Hamilton--Jacobi problem.
\qed

If $\alpha\in\df^1(Q)$ is a solution of the generalized
Hamiltonian Hamilton--Jacobi problem, taking into account
Corollary~\ref{XZHcor}, we can conclude that
the $\pi_Q$-projection of the integral curves of $Z_H$
contained in ${\rm Im}\,\alpha$ are
the integral curves of $X$.

Observe also that until now we have not used that $Z_H$ is a 
Hamiltonian vector field, so these results actually hold for every vector
field $Z\in\vectorfields{T^*Q}$.
When $Z_H$ is the Hamiltonian
vector field of a Hamiltonian system, the
above results can be expressed in terms of the Hamiltonian function.

As in the Lagrangian case, we can obtain an equation not
involving directly the dynamical vector field:

\begin{theorem} 
Given the Hamiltonian vector field $Z_H\in \vectorfields{T^*Q}$, 
a 1-form $\alpha\in\df^1(Q)$ is a solution of the generalized Hamiltonian
Hamilton--Jacobi problem if, and only if,
\begin{equation}
i(X)d\alpha=-d(\alpha^*H) \ ,
\label{cerobis}
\end{equation}
where $X=T\pi_Q\circ Z_H\circ\alpha$ is the vector field associated with
$\alpha$ by means of the Hamiltonian vector field $Z_H$.
\label{alphaX}
\end{theorem}
\proof
 From the Hamiltonian dynamical equation (\ref{Heq}) for $Z_H$ we obtain
$$
\alpha^*i(Z_H)\omega=\alpha^*dH=d(\alpha^*H) \ .
$$
Furthermore, $\theta$ is the canonical form of $T^*Q$,
so $\alpha^*\theta=\alpha$, and then
\begin{equation}
\alpha^*\omega = -\alpha^*d\theta=-d(\alpha^*\theta)=-d\alpha \ ,
\label{cero}
\end{equation}
therefore, as $X$ and $Z_H$ are $\alpha$-related, we have
$$
\alpha^*i(Z_H)\omega=i(X)\alpha^*\omega=-i(X)d\alpha
\,,
$$
which yields~(\ref{cerobis}). 

To prove the converse, first let us define
$$
D_H=Z_H\circ\alpha-T\alpha\circ X\colon Q\to TT^*Q 
\,,
$$
which is a vector field along $\alpha$. 
We have to prove that $D_H=0$.
First we have that $D_H$ is $\pi_Q$-vertical; in fact,
as $\pi_Q\circ\alpha={\rm Id}_Q$,
\beann
T\pi_Q\circ D_H &=& T\pi_Q\circ (Z_H\circ\alpha-T\alpha\circ X)=
T\pi_Q\circ (Z_H\circ\alpha-
T\alpha\circ T\pi_Q\circ Z_H\circ\alpha)
\\ &=&
T\pi_Q\circ Z_H\circ\alpha-T\pi_Q\circ Z_H\circ\alpha=0 \ .
\eeann
Furthermore, from the Hamiltonian dynamical equation (\ref{Heq}) 
and the hypothesis, as $\alpha^*\omega=-d\alpha$, we have
the following relations:
\beann
\alpha^*i(Z_H)\omega = \alpha^*dH &=& d(\alpha^*H)
\ ,
\\
i(X)\alpha^*\omega=-i(X)d\alpha &=& d(\alpha^*H) \ ,
\eeann
and hence $\alpha^*i(Z_H)\omega-i(X)\alpha^*\omega=0$.
Therefore, for every $q\in Q$ and $Y_q\in T_qQ$, we have
\beann
0 &=& (\alpha^*i(Z_H)\omega-i(X)\alpha^*\omega)_q(Y_q)=
\omega_{\alpha(q)}(Z_H(\alpha(q)),T_q\alpha(Y_q))-
\omega_{\alpha(q)}(T_q\alpha(X_q),T_q\alpha(Y_q))
\\
&=& \omega_{\alpha(q)}(D_H(q),T_q\alpha(Y_q)) \ .
\eeann
Moreover, as ${\rm V}(\pi_Q)$ (the $\pi_Q$-vertical subbundle of
$TT^*Q$) is a Lagrangian distribution in $(T^*Q,\omega)$,
for every $\pi_Q$-vertical
vector field $V\in\vectorfields{T^*Q}$ we have that
$$
\omega_{\alpha(q)}(D_H(q),V(\alpha(q)))=0 \ ;
\,,
$$
for every $q\in Q$. But $T_{\alpha(q)}T^*Q=
T_{\alpha(q)}({\rm Im}\,\alpha)\oplus{\rm V}_{\alpha(q)}(\pi_Q)$,
hence we have proved that
$$
\omega_{\alpha(q)}(D_H(q),Z(\alpha(q)))=0
$$
for every $q\in Q$ and $Z\in\vectorfields{T^*Q}$;
since $\omega$ is non-degenerate, we conclude that $D_H=0$,
or what is equivalent,
$X$ and $Z_H$ are $\alpha$-related, and thus
$\alpha$ is a solution of the generalized Hamiltonian
Hamilton--Jacobi 
\qed

As in the Lagrangian case,
in general, to solve the generalized Hamiltonian Hamilton--Jacobi problem
is a difficult task. So it is convenient to
consider the following less general problem,
which constitutes the standard version of the Hamiltonian Hamilton--Jacobi 
problem:

\paragraph{Hamiltonian Hamilton--Jacobi problem}
{\it
Given a vector field $Z_H\in \vectorfields{T^*Q}$,
the {\rm Hamiltonian Hamilton--Jacobi problem} consists in finding
a solution $\alpha\in\df^1(Q)$ of the generalized Hamiltonian Hamilton--Jacobi problem
which is moreover {\rm closed}, $d\alpha=0$.
}

\medskip

As a consequence, every point has an open neighbourhood
$U\subset Q$, where there is a function $W\in\cinfty{U}$ such that $\alpha=dW$.

Notice also that, because of (\ref{cero}), 
the closeness condition 
$d\alpha=0$ is equivalent to $\alpha^*\omega=0$.

\medskip

A straightforward consequence of the previous theorem is:

\begin{corollary}
A closed 1-form $\alpha$ is a solution of the Hamiltonian
Hamilton--Jacobi problem if, and only if,
$d(\alpha^*H)=0$.
\label{coro1}
\qed
\end{corollary}

Observe that, if $\alpha\in\df^1(Q)$ is a solution of the Hamiltonian
Hamilton--Jacobi problem, as $\alpha^*\omega=0$,
then ${\rm Im}\,\alpha$ is a Lagrangian submanifold
of $(T^*Q,\omega)$, contained in a level set of
$H$, because (\ref{cerobis}) implies that $d(\alpha^*H)=0$.
In fact, $\dim\,{\rm Im}\,\alpha=n$ and,
if $j\colon{\rm Im}\,\alpha\hookrightarrow T^*Q$
denotes the natural embedding, we have that $j^*\omega=0$.
Thus we recover some geometrical aspects of
the classical Hamiltonian Hamilton--Jacobi theory.

We can summarize the above results in the following:

\begin{proposition}
Let $\alpha\in\df^1(Q)$ be a closed 1-form.
Then, the following assertions are equivalent:
\begin{enumerate}
\item 
$\alpha$ is a solution of the
Hamiltonian Hamilton--Jacobi problem.
\item
$d(\alpha^*H)=0$.
\item
${\rm Im}\,\alpha$ is a Lagrangian submanifold of 
$T^*Q$ invariant by $Z_H$.
\item
The integral curves of $Z_H$ with initial conditions
in ${\rm Im}\,\alpha$ project onto the integral curves of 
$X=T\pi_Q\circ Z_H\circ\alpha$.
\end{enumerate} 
If moreover $\alpha=dW$, then these conditions can also be written as
\begin{enumerate}
\setcounter{enumi}{4}
\item $H\circ dW$ is locally constant.
\qed
\end{enumerate}
\label{Hamconverse}
\end{proposition}

\paragraph{Coordinate expressions}

Let us see how all the objects presented appear
when we consider coordinates $(q^i)$ on~$Q$, 
and the corresponding natural coordinates 
$(q^i,\dot q^i)$ and $(q^i,p_i)$ 
on its tangent and cotangent bundles.

First, the coordinate expression of the Hamiltonian vector field 
$Z_H$ is given by 
$$
Z_H(q,p) = (q,p; \partial H/\partial p, -\partial H/\partial q) .
$$

Consider a 1-form $\alpha \in \df^1(Q)$ 
and a vector field $X \in \vectorfields{Q}$.
In coordinates they read
$\alpha = a_i \,\d q^i$ and
$X = X^i \,\partial/\partial q^i$.
Then $\Tan \alpha \circ X$ and $Z_H \circ \alpha$ 
are vector fields along $\alpha$, which read
\begin{align*}
&(\Tan \alpha \circ X)(q) = 
\left(
q^i,a_i(q); X^i(q),\frac{\partial a_i}{\partial q^j} X^j(q)
\right) ,
\\
&(Z_H \circ \alpha) (q) = 
\left(
q^i,a_i(q); 
\frac{\partial H}{\partial p_i}(q,a(q)), 
-\frac{\partial H}{\partial q^i}(q,a(q))
\right) .
\end{align*}
Therefore their difference is
$$
(\Tan \alpha \circ X - Z_H \circ \alpha)(q) = 
\left(
q^i,a_i(q); 
X^i(q) - \frac{\partial H}{\partial p_i}(q,a(q)),
\frac{\partial a_i}{\partial q^j} X^j(q) + 
\frac{\partial H}{\partial q^i}(q,a(q))
\right) .
$$
Its vanishing determines $X = \FH \circ \alpha$, that is: 
\beq
X^i(q) = \frac{\partial H}{\partial p_i}(q,a(q)) .
\eeq
With this $X$, the preceding difference becomes
\beq
(\Tan \alpha \circ X - Z_H \circ \alpha)(q) = 
\left(
q^i,a_i(q); 
0,
\frac{\partial a_i}{\partial q^j}(q) 
\frac{\partial H}{\partial p_j}(q,a(q)) + 
\frac{\partial H}{\partial q^i}(q,a(q))
\right) .
\label{hj1}
\eeq
So, the condition for $\alpha$ to be a solution of the
generalized Hamiltonian Hamilton--Jacobi problem is
$$
\frac{\partial a_i}{\partial q^j}(q) 
\frac{\partial H}{\partial p_j}(q,a(q)) + 
\frac{\partial H}{\partial q^i}(q,a(q))
= 0 .
$$

Now let us consider 
$\ds
\d \alpha = \frac{\partial a_i}{\partial q^j} \d q^j \wedge \d q^i
$.
Then 
$\ds
i(X) \d \alpha = 
X^j \left( 
\frac{\partial a_i}{\partial q^j} - \frac{\partial a_j}{\partial q^i}
\right) \d q^i
$.
On the other hand, 
$\ds
\d H = 
\frac{\partial H}{\partial q^i} \d q^i + 
\frac{\partial H}{\partial p_i} \d p_i
$, 
so 
$\ds
\alpha^*(\d H) = 
\left(
\frac{\partial H}{\partial q^i}(q,a(q)) + 
\frac{\partial H}{\partial p_j} \frac{\partial a_j}{\partial q^i}(q,a(q)) 
\right) \d q^i
$.
Therefore we have
$$
i(X) \d \alpha + \alpha^*(\d H) =
\left.\left( 
X^j \frac{\partial a_i}{\partial q^j} 
+ 
(\frac{\partial H}{\partial p_j} -X^j) \frac{\partial a_j}{\partial q^i}
+
\frac{\partial H}{\partial q^i}
\right)\right|_{p=a(q)}
\d q^i .
$$
Again, with $X$ given as $\FH \circ \alpha$, 
this expression becomes
\beq
i(X) \d \alpha + \alpha^*(\d H) =
\left.\left( 
\frac{\partial H}{\partial p_j} \frac{\partial a_i}{\partial q^j} 
+ 
\frac{\partial H}{\partial q^i}
\right)\right|_{p=a(q)}
\d q^i .
\label{hj2}
\eeq

Finally, if $\alpha = \d W$, then $a_i = \partial W/\partial q^i$, 
and the last condition in Proposition~\ref{Hamconverse}
reads
$$
H(q^i,\partial W/\partial q^i) = \mathrm{const} ,
$$
which is the classical form of the 
time-independent Hamilton--Jacobi equation.

A careful look at the local expressions (\ref{hj1}) and~(\ref{hj2})
gives an alternative understanding of Theorem~\ref{alphaX}.
Using the vertical lift map again, we have:
\begin{proposition}
Let $\vl$ be the vertical lift map
of the cotangent bundle $\Tan^*Q$.
Given a 1-form $\alpha$ on~$Q$,
and the vector field $X = \FH \circ \alpha$ on~$Q$, 
we have
\beq
\vl \left(\strut \alpha,i(X) \d \alpha + \alpha^*(\d H) \right) 
=
\Tan \alpha \circ X - Z_H \circ \alpha .
\eeq
Therefore, 
$\Tan \alpha \circ X - Z_H \circ \alpha$ vanishes if, and only if,
$i(X) \d \alpha + \alpha^*(\d H)$ also does.
\qed
\end{proposition}

\subsection{Complete solutions}

As in the Lagrangian case,
we are interested in finding not only a particular solution
as described in the preceding section, but a complete solution
to the problem. In this way, we define:

\begin{definition}
Consider a solution $\alpha_\lambda$ depending on $n$ additional parameters
$\lambda\in\Lambda$, where $\Lambda\subseteq\R^n$ is an open set,
and suppose that the map $\Phi\colon Q\times\Lambda\to T^*Q$
given by $\Phi(q,\lambda)=\alpha_\lambda(q)$ is a local diffeomorphism. 
In this case $\{ \alpha_\lambda; \lambda\in\Lambda\}$ 
is said to be a {\rm complete solution} of the generalized
Hamiltonian Hamilton--Jacobi problem.
\end{definition}

 From the definition it follows that a complete solution provides $T^*Q$
with a foliation transverse to the fibers, and that the Hamiltonian vector field
$Z_H$ is tangent to the leaves. 

If $\{ \alpha_\lambda;\lambda\in\Lambda\}$ is a complete solution,
the integral curves of the vector fields 
provide all the integral curves of the
Hamiltonian vector field $Z_H$.
This means that, if $(q_0,p_0)\in{\rm Im}\,\alpha$,
then there is $\lambda_0\in\Lambda$ such that
$\alpha_{\lambda_0}(q_0)=p_0$, and the integral curve of
$X_{\lambda_0}$ through $q_0$, lifted
by $\alpha_{\lambda_0}$ to $T^*Q$, gives the integral
curve of $Z_H$ through $(q_0,p_0)$.
This justifies the name of ``complete solution''.

Furthermore, different transversal foliations of $(T^*Q,\omega)$,
with $Z_H$ tangent to the leaves, are different ways to collect
integral curves of $Z_H$ smoothly and such that
they project onto $Q$ in a coherent way:
integral curves of $Z_H$ in ${\rm Im}\,\alpha_\lambda$
project onto integral curves of the associated vector field~$X_\lambda$).

The relation between $Z_H$ and complete solutions is the same
as in the Lagrangian case, using first integrals
of $Z_H$ and the vector fields
$X_\lambda$ associated to $\alpha_\lambda$, for $\lambda\in\Lambda$.

Finally, if the foliation is Lagrangian in $(T^*Q,\omega)$,
then we have a complete solution of the Hamiltonian Hamilton--Jacobi problem.
In this case the above family of first integrals
are in involution.

\subsection{Equivalence between the Lagrangian and Hamiltonian formulations}
\protect\label{equi}

This section is devoted to the equivalence
between the Lagrangian and Hamiltonian Hamilton--Jacobi
theory. We have the following:

\begin{theorem}
Let $(TQ,\omega_L,E_L)$ be a hyper-regular Lagrangian system, and
$(T^*Q,\omega,H)$ its associated Hamiltonian system.
Then there exists a bijection between the set of solutions of the
(generalized) Lagrangian Hamilton--Jacobi problem and
the set of solutions of the (generalized) Hamiltonian Hamilton--Jacobi problem.
This bijection is given by composition with the Legendre map:
$X\mapsto \alpha=\FL\circ X$.
\label{equivHJ} 
\end{theorem} 
\proof
Suppose $X\in\vectorfields{Q}$ satisfies
$TX\circ X=\Gamma_L\circ X$, and
$T\FL\circ\Gamma_L=Z_H\circ\FL$.
Let $\alpha=\FL\circ X$, then
$$
T\alpha\circ X=T(\FL\circ X)=T\FL\circ TX\circ X=
T\FL\circ\Gamma_L\circ X=Z_H\circ\FL\circ X=
Z_H\circ\alpha\ ,
$$
hence $\alpha$ is a solution of the Hamiltonian problem.

Furthermore, if $E_L\circ X=const.$, and $\alpha=\FL\circ X$,
then $E_L\circ\FH\circ\alpha=const.$,
that is, $H\circ\alpha=const.$

Conversely, suppose $\alpha\in\Omega^1(Q)$
satisfies $T\alpha\circ X=Z_H\circ\alpha$, and
$\Gamma_L\circ\FH=T\FH\circ Z_H$,
where $\FH$ denotes the fiber derivative of the Hamiltonian,
which satisfies $\FH=(\FL)^{-1}$, because the system is hyper-regular.
Let $X=\FH\circ\alpha$, then
$$
TX\circ X=T(\FH\circ\alpha)\circ X=
T\FH\circ T\alpha\circ X=
T\FH\circ Z_H\circ\alpha=
\Gamma_L\circ\FH\circ\alpha=
\Gamma_L\circ X \ ,
$$
hence $X$ is a solution of the corresponding Lagrangian problem.

In addition, 
if $H\circ\alpha=const.$, and $X=\FH\circ\alpha$,
then $H\circ\FL\circ X=const.$, that is,
$E_L\circ X=const.$
\qed

This result can be extended to complete solutions in a natural way.

As is obvious, for regular but non hyper-regular Lagrangians,
all this holds only in the local open sets where $\FL$
is a diffeomorphism.

\subsection{Alternative Lagrangian descriptions}
\protect\label{altlag}

Let us consider now the case of a regular system
admitting alternative Lagrangian descriptions;
that is, suppose there are regular Lagrangians functions
$L,L'\in\cinfty{TQ}$, $L\not= L'$,
giving rise to the same dynamical vector field
$\Gamma_{L}=\Gamma_{L'}\in\vectorfields{TQ}$
solution of the equations
$$
i(\Gamma)\omega_{L}=dE_{L}\,, \quad i(\Gamma)\omega_{L'}=dE_{L'} \ .
$$
So, we have the same dynamics, but two different
symplectic structures.

\begin{itemize}
\item 
If $X$ is a solution of the generalized Hamilton--Jacobi problem for one of the Lagrangians,
then it is also a solution of the generalized problem for the second Lagrangian.
A similar result does not hold for the non generalized problem, that is, a solution
$X$ of the Hamilton--Jacobi problem for one of the Lagrangians will not be (in general)
a solution of  the Hamilton--Jacobi problem for the other Lagrangian,
that is, $X^*\omega_L=0$, but $X^*\omega_{L'}\neq0$.
\item
It is natural to compare with the situation in the
Hamiltonian formalism, that is, in $T^*Q$.
Instead of $\omega_{L}$ and $\omega_{L'}$,
we have a symplectic structure $\omega\in\df^2(T^*Q)$,
but different Hamiltonian functions $H, H' \in\cinfty{T^*Q}$.
Thus the same solution $X$ of the Lagrangian Hamilton--Jacobi problem
leads to two solutions $\alpha$ and $\alpha'$ of the
Hamiltonian Hamilton--Jacobi problem corresponding to
the Hamiltonians $H$ and $H'$ respectively.
Nevertheless, notice that
$$
\frac{dq^i}{dt}=\derpar{H}{p_i}(q,p)\Big\vert_{p=\derpar{W}{q}}=
\derpar{H'}{p_i}(q,p)\Big\vert_{p=\derpar{W'}{q}}
$$
where $\alpha=d W$ and $\alpha'=d W'$ locally.
These are the equations for the integral curves of $X$.
However, the corresponding dynamical vector fields on $T^*Q$ related to $X$ are different.
In other words, the difference between $Z_H\circ\alpha$
and $Z_{H'}\circ\alpha'$ is a vertical vector field, which in general does not vanish.
\item
The case of dynamical systems described by two alternative equivalent
Lagrangians motivates the use of two different symplectic
structures in $T^*Q$ in the following way.
Let $L$ and $L'$ be equivalent hyper-regular Lagrangians,
and $\omega_L$ and $\omega_{L'}$ their corresponding 
Lagrangian $2$-forms. If $\omega_0\in\df^2(T^*Q)$
is the canonical $2$-form in $T^*Q$, we have that
$\FL^*\omega_0=\omega_L$. Then, let $\omega_1\in\df^2(T^*Q)$
be another symplectic structure in $T^*Q$ such that
$\FL^*\omega_1=\omega_{L'}$. Hence,
$\FL'\circ\FL^{-1}$ is  a base-preserving transformation from
$(T^*Q,\omega_0)$ to $(T^*Q,\omega_1)$.
These transformations have been called fouling transformations~\cite{MMS83}.
\end{itemize}
As an example, it is not difficult to show this construction
for the two-dimensional harmonic oscillator
described both by
$L=\frac{1}{2}((v^1)^2+(v^2)^2-(q^1)^2-(q^2)^2)$, and
$L'=v^1v^2-q^1q^2$
(see example~2 in Section~\ref{cs}).

These considerations show that, in order to incorporate
alternative Lagrangian or Hamiltonian descriptions in the Hamilton--Jacobi setting,
we must introduce generalized solutions.

\section{
The Hamilton--Jacobi problem for unconstrained singular Lagrangian systems}
\protect\label{uncons}

In this section we are going to show how our procedure can be extended
to singular Lagrangians without secondary constraints.

\subsection{Lagrangian formulation}
\protect\label{LRHJsing2}

Now we consider a singular Lagrangian $L\in\cinfty{TQ}$.
We recall that the Euler--Lagrange equation for a \textsc{sode} $\Gamma$ is the equation
\begin{equation}
\label{ELeq-fields}
i(\Gamma)\omega_L=dE_L.
\end{equation}
A curve $\map{\gamma}{\Real}{Q}$ is a solution of 
the Euler--Lagrange equation if $\xi=\dot{\gamma}$ satisfies
\begin{equation}
\label{ELeq-curves}
i(\dot{\xi})\omega_L=dE_L\circ\xi.
\end{equation}

We shall only consider the case of singular Lagrangians for which
the following assumption holds:

\begin{assumption} 
The Lagrangian dynamical equation (\ref{ELeq-fields})
has a \textsc{sode} solution $\Gamma\in\vectorfields{TQ}$ everywhere defined in $TQ$ and the rank of $T\FL$ is constant.
\label{unconslag2}
\end{assumption} 

The constancy of the rank of $T\FL$ is equivalent to saying that
$\omega_L$ has also constant rank, hence $\omega_L$ is a presymplectic form.

Under this assumption, the set of \sode\ solution vector fields 
is the set of sections of an affine bundle $\cala\to TQ$, 
modeled on the vector bundle 
$\Ker T\FL\to TQ$. More precisely, the fiber of $\cala$ at $v\in TQ$ is 
\[
\cala_v=\set{V\in T_v(TQ)}{T\tau_Q(V)=v\quand
i(V)\omega_L\vert_v=dE_L\vert_v}.
\]
A curve $\map{\gamma}{\Real}{Q}$ 
is a solution of the Euler--Lagrange equations if, and only,
the curve $\xi=\dot{\gamma}$ satisfies $\dot{\xi}(t)\in\cala_{\xi(t)}$,
for every $t\in\Real$.
Observe that if a vector $V\in T_v(TQ)$ satisfies $T\tau_Q(V)=v$
then the linear 1-form  $i(V)\omega_L\vert_v-dE_L\vert_v$ is semibasic.

The generalized Lagrangian Hamilton--Jacobi problem
for these kinds of Lagrangians can be stated as follows:

\paragraph{Generalized Lagrangian Hamilton--Jacobi problem for
unconstrained singular Lagrangians}
{\it To find a vector field $\map{X}{Q}{TQ}$ such that,
if $\map{\gamma}{\Real}{Q}$ is an integral curve
of $X$, then $\map{\xi=X\circ\gamma}{I}{TQ}$ 
is a solution of the Euler--Lagrange equation (\ref{ELeq-curves})}.

\medskip

In a similar way to the first part of Section~\ref{statementoftheproblem},
we have the following result:

\begin{theorem}
The following conditions for a vector field $X\in\vectorfields{Q}$ are equivalent:
\begin{enumerate} 
\item $X$ is a solution of the generalized Lagrangian Hamilton--Jacobi problem. 
\item $X$ satisfies the condition 
${\rm Im}\,(TX\circ X)\subset\cala|_{{\rm Im}\, X}$.
\item $X$ satisfies the equation $i(X)(X^*\omega_L)=d(X^*E_L)$.
\item For every $v\in{\rm Im}\, X$ there exists $w\in\cala_v$ 
such that $w$ is tangent to ${\rm Im}\, X$.
\item The submanifold ${\rm Im}\, X$ is such that for every
 initial condition $v\in{\rm Im}\, X$ there is a solution of 
the Euler--Lagrange equations which is entirely contained in
 the submanifold ${\rm Im}\, X$.
\end{enumerate}
\end{theorem} 
\begin{proof}

\noindent [$(1) \Rightarrow (2)$] 
Let $q\in Q$ be arbitrary, and consider the integral curve $\gamma$ of $X$ 
such that $\gamma(0)=q$. Denote by $\xi=\dot{\gamma}$ 
the tangent lift of $\gamma$. Since condition $(1)$ is satisfied
and $\gamma$ is an integral curve of $X$,
we have that $\dot{\xi}(t)\in\cala_{\xi(t)}$. 
But since $\xi=X\circ\gamma$ we have 
$\dot{\xi}=TX\circ\dot{\gamma}=TX\circ X\circ\gamma$, 
which at $t=0$ gives
 $TX(X(q))=\dot{\xi}(0)\in\cala_{\xi(0)}=\cala_{X(q)}$.

\noindent [$(2) \Rightarrow (3)$]
Assume that ${\rm Im}\,(TX\circ X)\subset\cala|_{{\rm Im}\, X}$, 
so that for every $q\in Q$ we have that $T_qX(X(q))\in\cala_{X(q)}$. 
Then, from the definition of $\cala$ 
we have that $\omega_L(T_qX(X(q)),W)=\pai{dE_L|_{X(q)}}{W}$
 for every $W\in T_{X(q)}(TQ)$.
In particular, if we take $W=T_qX(w)$ for arbitrary $w\in T_qQ$,
 we have that $\omega_L(T_qX(X(q)),T_qX(w))=\pai{dE_L|_{X(q)}}{T_qX(w)}$,
 or in other words 
$(X^*\omega_L)_q(X(q),w)=\pai{X^*(dE_L)|_q}{w}$. 
Since this equality holds for every $w\in T_qQ$ and every $q\in Q$,
we deduce that $i(X)(X^*\omega_L)=X^*dE_L$.

\noindent [$(3) \Rightarrow (4)$]
Let $X$ be a vector field satisfying condition (3) and
$v\in\mathrm{Im}\,X$, so that $v=X(q)$ for $q=\tau_Q(v)$.
The vector $w=TX(v)$ satisfies the required properties.
Indeed, on one hand it is clear that $w$ is tangent to the image of $X$,
and on the other we have that
$T\tau_Q(w)=T\tau_Q(TX(v))=T(\tau_Q\circ X)(v)=v$,
so that we have to prove that that linear 1-form $i(w)\omega_L|_v-dE_L|_v$ vanishes.
Since such 1-form is semibasic, we just need to prove that it vanishes
when applied to elements of the form $TX(u)$ for $u\in T_qQ$:
\begin{align*}
(i(w)\omega_L|_v-dE_L|_v)(TX(u))
&=\omega_L|_{X(q)}(TX(v),TX(u))-dE_L|_{X(q)}(TX(u))\\
&=(X^*\omega_L)_q(v,u)-d(X^*E_L)_q(u)\\
&=(X^*\omega_L)_q(X(q),u)-d(X^*E_L)_q(u),
\end{align*}
which vanishes in view of the condition $i(X)X^*\omega_L-d(X^*E_L)=0$.

\noindent [$(4) \Rightarrow (1)$]
Assume that for every element $v\in{\rm Im}\, X$ there exists $w\in\cala_v$,
which is tangent to ${\rm Im}\, X$. In other words, 
for every $q\in Q$ (and hence $v=X(q)$) there exists $w\in\cala_{X(q)}$ 
such that $w=T_qX(z)$ for some $z\in T_qQ$. 
But the first condition for the element $w$ to be in $\cala$ 
is $T\tau_Q(w)=\tau_{TQ}(w)$, which for $w=T_qX(z)$ is just $z=X(q)$. 
Therefore, the vector $w$ is $w=T_qX(X(q))$ and it is $\cala_{X(q)}$. 
Since this is true for every $q\in Q$, we have proved that 
${\rm Im}\, (TX\circ X)\subset \cala|_{{\rm Im}\, X}$, 
which was shown to be equivalent to condition~(1).

Finally, $(4)$ and $(5)$ are clearly equivalent, 
and both are equivalent to the integrability of the restriction of $\cala$ to 
${\rm Im}\, X$ (see the remark below this proof). 
\qed
\end{proof}

{\bf Remark:}
Let us recall a few facts from the theory of implicit differential systems, 
in particular, when the implicit system is just an affine subbundle 
of the tangent bundle.
Let $\cala\to M$ be an affine subbundle of $TM$ and let $N$ be 
a submanifold of $M$. Consider the restriction $\cala|_N$ 
of the subbundle $\cala$ to~$N$. The following properties are equivalent:
\begin{enumerate}
\item The restriction of $\cala$ to $N$ satisfies the integrability condition
for implicit differential equations \cite{GP93,MMT95}.
\item 
For every initial condition $m\in N$ there exists a curve solution 
of the system which is entirely contained in $N$.
\item 
For every $m\in M$ there exists $w\in\cala_m$ such that $w$ is tangent to $N$.
\end{enumerate}
Roughly speaking, the proofs of these facts are as follows: 
$(1)$ and $(2)$ are equivalent by definition 
of an integrable implicit differential system.
[(2)$\Rightarrow$(3)] is obvious: given $m\in M$
 take the solution $\gamma(t)$ passing through $m$ and contained in $M$, and then
 $w=\dot{\gamma}(0)$ is tangent to $M$.
 [(3)$\Rightarrow$(2)] Take a local section of $\cala\cap TN$,
 and an integral curve of such a section is a curve contained in $M$.

\medskip

As in the regular case, we can state the following particular problem:

\paragraph{Lagrangian Hamilton--Jacobi problem for unconstrained singular Lagrangians}
{\it To find solutions $X$ to the generalized
Lagrangian Hamilton--Jacobi problem for unconstrained singular Lagrangians
satisfying $X^*\omega_L=0$}.

\medskip

The main results for this situation are summarized in the following:

\begin{proposition}
The following assertions for a vector field $X\in\vectorfields{Q}$ are equivalent:
\begin{enumerate}
\item 
$X$ is a solution of the Lagrangian Hamilton--Jacobi problem.
\item
${\rm Im}\, X$ is an isotropic submanifold of $(TQ,\omega_L)$ and
${\rm Im}\,(TX\circ X)\subset\cala|_{{\rm Im}\, X}$.
\item
$d(X^*\theta_L)=0$ and $d(X^*E_L)=0$.
\item
${\rm Im}\, X$ is an isotropic submanifold of $(TQ,\omega_L)$ and for every
$v\in{\rm Im}\, X$ there exists $w\in\cala_v$ such that $w$ is tangent to ${\rm Im}\, X$.
\item
${\rm Im}\, X$ is an isotropic submanifold of $(TQ,\omega_L)$,
and for every initial condition in ${\rm Im}\, X$
there exists a solution of the Euler--Lagrange equations entirely contained in ${\rm Im}\, X$.
\end{enumerate} 
\end{proposition}
\begin{proof}
They are consequences of the last theorem, taking into account that
${\rm Im}\, X$ is isotropic if, and only if, $X^*\omega_L=0$,
and this is equivalent to $d(X^*\theta_L)=0$.
\qed
\end{proof}

\subsection{Hamiltonian formulation}

When the Lagrangian is singular, in general,
there is no satisfactory Hamiltonian formalism
unless certain regularity conditions hold. We will assume in what follows that:

\begin{assumption}
The Lagrangian $L$ is \emph{almost-regular}, that is:
$P=\FL(TQ)$ is a closed submanifold of $T^*Q$, $\FL$ is a submersion onto its image $P$,
and the fibers $\FL^{-1}(\FL(p))$, for every $p\in TQ$,  are connected submanifolds of $TQ$. 
\end{assumption}

The natural embedding of $P$ into $T^*Q$ will be denoted 
$\jmath_0\colon P\hookrightarrow T^*Q$. 
Denote by $\FL^0$ the map $\FL^0\colon TQ\to P$ defined by the relation
$\jmath_0\circ\FL^0=\FL$.

For an almost-regular Lagrangian system $(TQ,L)$ there exists a Hamiltonian formalism. 
The associated Hamiltonian system is $(P,\omega_0, H_0)$,
 where $\omega_0=\jmath_0^*\omega$ is a presymplectic form,
 and $H_0\in\cinfty{P}$ is the Hamiltonian function,
 defined by the equation $\FL^{0*}H_0=E_L$.

The Hamilton equation is the presymplectic equation
\begin{equation}
i(Z)\omega_0=dH_0,
\label{Heq-fields}
\end{equation}
for a vector field $Z\in\vectorfields{P}$.
 Under our assumptions this equation
 has solution everywhere in $P$, although it is not unique~\cite{Ca-90,K2,K1}.
 The set of solutions is the set of sections of an affine subbundle
 $\calb\to P$ of $T(T^*Q)$, modeled on the vector subbundle 
$\Ker(\omega_0)\to P$. The fiber over a point $\alpha\in P$ is
\[
\calb_\alpha=\{V\in T_\alpha(T^*Q)\mid i(V){\omega_0}_{|\alpha}={dH_0}_{_\alpha}\}.
\]
A curve $\map{\mu}{\Real}{T^*Q}$ is a solution of the Hamilton equations 
if it satisfies
\begin{equation}
i(\dot{\mu})\,\omega_0=dH_0\circ\mu.
\label{Heq-curves}
\end{equation}
Hence, the curve $\mu$ is a solution of the Hamilton equation
if, and only if, $\dot{\mu}(t)\in\calb_{\mu(t)}$.

Bearing in mind the above comments and the results for the Lagrangian and 
the Hamiltonian regular cases,
the generalized version of the Hamiltonian Hamilton--Jacobi problem
for these kinds of singular systems can be stated in the following way,
which is not exactly as in the regular case:

\paragraph{Generalized Hamiltonian Hamilton--Jacobi problem for
unconstrained singular Lagrangians}
{\it To find vector fields $\map{X}{Q}{TQ}$ such that, if
$\map{\gamma}{\Real}{Q}$ is an integral curve
of $X$ then $\mu=\FL^0\circ X\circ\gamma$ is a curve 
solution of the Hamilton equation~(\ref{Heq-curves}).

Observe that $\map{\alpha=\FL^0\circ X}{Q}{P}$
 is a section of the projection $\map{\pi_Q^0=\pi_Q\circ\jmath_0}{P}{Q}$.
 We will say that $\alpha$ is the $1$-form associated with the 
particular chosen solution $X$.}

\medskip

In this way, all the definitions, results and comments
stated in Section \ref{Hamsec} 
hold for the manifold $P$ instead of $T^*Q$.
In particular:

\begin{theorem}
The following conditions for a vector field $X\in\vectorfields{Q}$ are equivalent
\begin{enumerate} 
\item $X$ is a solution of the generalized 
Hamiltonian Hamilton--Jacobi problem for the unconstrained singular Lagrangian $L$,
 with associated 1-form $\alpha$. 
\item $X$ satisfies the condition
 ${\rm Im}\,(T\alpha\circ X)\subset\calb|_{{\rm Im}\, \alpha}$
\item $X$ satisfies the equation $i(X)(\alpha^*\omega_0)=d(\alpha^*H_0)$.
\item For every $\lambda\in{\rm Im}\, \alpha$ there exists
 $w\in\calb_\lambda$ such that $w$ is tangent to ${\rm Im}\,\alpha$.
\item The submanifold ${\rm Im}\, \alpha$ is such that,
 for every initial condition in ${\rm Im}\, \alpha$
 there is a curve solution of the Hamilton equations
 which is entirely contained in the submanifold ${\rm Im}\, \alpha$.
 \qed
\end{enumerate}
\end{theorem}

As above, we can state the particular case:

\paragraph{Hamiltonian Hamilton--Jacobi problem for unconstrained singular Lagrangians}
{\it To find solutions $\alpha$ of the generalized singular
Hamiltonian Hamilton--Jacobi problem for unconstrained singular Lagrangians
satisfying $\alpha^*\omega_0=0$.}

\medskip

And we obtain:

\begin{proposition}
The following assertions for a 1-form $\map{\alpha}{Q}{P\subset T^*Q}$
 are equivalent:
\begin{enumerate}
\item 
$\alpha$ is a solution of the Hamiltonian Hamilton--Jacobi problem
for the unconstrained singular Lagrangian $L$.
\item
${\rm Im}\,\alpha$ is an isotropic submanifold of
 $P,\omega_0)$ and ${\rm Im}\,(T\alpha\circ X)\subset\calb|_{{\rm Im}\,\alpha}$.
\item
$d(j_0\circ\alpha)=0$ and $d(\alpha^*H)=0$.
\item
${\rm Im}\, \alpha$ is an isotropic submanifold of $(P,\omega_0)$ and,
 for every $\lambda\in{\rm Im}\,\alpha$,
 there exists $w\in\calb_\lambda$ such that $w$ is tangent to ${\rm Im}\,\alpha$.
\item
${\rm Im}\, \alpha$ is an isotropic submanifold of $(P,\omega_0)$,
 and for every initial condition in ${\rm Im}\, \alpha$
 there exists a curve solution of the Hamilton equations
 entirely contained in ${\rm Im}\,\alpha$.
 \qed
\end{enumerate} 
\end{proposition}

As a final remark, the equivalence
between the Lagrangian and Hamiltonian Hamilton--Jacobi
problem in the unconstrained singular case
is straightforward, taking into account how the
problem has been stated in the Hamiltonian formalism.

\section{The Hamilton--Jacobi problem for time-dependent regular systems}
\protect\label{td}

\subsection{The extended homogeneous Lagrangian formalism}

The geometric formalism for non-autonomous Lagrangian and
Hamiltonian systems exhibits some differences with respect to the autonomous formalism
(see \cite{CrPrTh,CraMaSa} for the details). 
In the non-relativistic Lagrangian formalism of time-dependent dynamical systems,
the configuration space is a bundle 
$\pi\colon E\to\R$, known as the configuration bundle,
and the velocity-phase space is the first-order jet bundle 
$\pi^1\colon J^1\pi\to E$. 
Given fibered coordinates $(t,q^i)$ on $E$, we get 
fibered coordinates $(t,q^i,v^i)$ on $J^1\pi$. 

A non-autonomous Lagrangian is a function $L\in\cinfty{J^1\pi}$. 
In this case, the associated Cartan 1-form $\Theta_L$ is 
the 1-form on $J^1\pi$ whose coordinate expression is
\[
\Theta_L=\pd{L}{v^i}(dq^i-v^idt)+Ldt.
\]
The Lagrangian is regular if the dimension of the kernel of the
Cartan 2-form $\Omega_L=-d\Theta_L$ is 1. 
In this case a unique vector field $\Gamma$, the dynamical vector field, 
is determined by the dynamical equation $i(\Gamma)\Omega_L=0$, 
together with the normalization condition $i(\Gamma)dt=1$, 
which ensures that integral curves of $\Gamma$ 
are parametrized by the time coordinate $t$.

It follows from the above description that our 
Hamilton--Jacobi theory does not apply directly to time-dependent systems. 
A way to solve this problem is to describe non-autonomous systems 
by the so-called homogeneous formalism (see~\cite{Klein},
for a friendly introduction see Section 2.3.1 in \cite{EMS04}),
as we are about to explain. Instead of the first jet bundle 
$J^1\pi$, we consider the tangent bundle $\tau_E\colon TE\to E$, 
which is called the extended Lagrangian phase space,
and we will define a new Lagrangian in this extended space 
whose solutions are related to the solutions of the original system.
 Natural coordinates on $TE$ will be denoted by $(x^0,x^i,w^0,w^i)$. 

The manifold $J^1\pi$ can be canonically embedded into $TE$ 
by means of the map $i\colon J^1\pi\to TE$ given by 
$i(j^1_{t_0}\sigma)=\dot{\sigma}(t_0)$. 
In fact, the image of $i$ is included into the open submanifold 
$\widehat{TE}\subset TE$ of vectors which are not vertical over $\R$. 
Conversely, we can define a map $p\colon\widehat{TE}\to{J^1\pi}$, 
which is a left inverse of $i$, defined as follows:
for $w=\dot{\gamma}(0)\in\widehat{TE}$, we consider the function 
$\varphi=\pi\circ\gamma\colon \R\to\R$;
this function is locally invertible in a neighborhood of $s=0$ since 
$\dot{\varphi}(0)\neq0$, thus we can consider 
$\sigma=\gamma\circ\varphi^{-1}$, which is a local section of $\pi$. 
The 1-jet of $\sigma$ at the point $t_0=\varphi(0)$ is well defined 
(it does not depend on the choice of the curve 
$\gamma$ that represents $w$), and we define $p(w)=j^1_{t_0}\sigma$. 
In coordinates, the expression of $i$ and $p$ are 
\[
i(t,q^i,v^i)=(t,q^i,1,v^i)
\qquad\text{and}\qquad
p(x^0,x^i,w^0,w^i)=\left(x^0,x^i,\frac{w^i}{w^0}\right).
\]
 From the above expression it is clear that 
$p\circ i=\operatorname{id}$ (but $i\circ p\neq\operatorname{id}$), 
and it is easy to see that the kernel of 
$Tp$ is generated by the Liouville vector field $\Delta$ on $TE$ restricted to $\widehat{TE}$.

Next we define a Lagrangian function $\hat{L}\in\cinfty{\widehat{TE}}$ 
in such a way that the action defined by a curve in the jet formalism 
and the action defined by the corresponding curve in
the extended formalism coincide, that is, 
with the same notation as above we look for a function $\hat{L}$ such that
\[
\int_{t_0}^{t_1}(j^1\sigma)^*L\,dt=\int_{s_0}^{s_1}\dot{\gamma}^*\hat{L}\,ds
\]
under the change of variable $t=\varphi(s)$. 
It follows that the Lagrangian $\hat{L}$ is defined by
\[
\hat{L}(\dot{\gamma}(0))=L(j^1_{t_0}\sigma)\dot{\varphi}(0).
\]
In other words $\hat{L}=w^0(p^*L)$, which in coordinates reads 
\[
\hat{L}(x^0,x^i,w^0,w^i)=L(x^0,x^i,w^i/w^0)w^0
\]
which is homogeneous of degree $1$.

\begin{proposition}
The following relations hold:
\begin{enumerate}
\item $\theta_{\hat{L}}=p^*\Theta_L$ and $\Theta_L=i^*\theta_{\hat{L}}$.
\item $\omega_{\hat{L}}=p^*\Omega_L$ and $\Omega_L=i^*\omega_{\hat{L}}$.
\item $E_{\hat{L}}=0$.
\end{enumerate}
\end{proposition}
\proof
A simple calculation in coordinates shows that
\[
\theta_{\hat{L}}=
\pd{\hat{L}}{w^0}dx^0+\pd{\hat{L}}{w^i}{dx^i}
=-p^*E_Ldx^0+p^*\left(\pd{L}{v^i}\right)dx^i
=p^*\left(-E_Ldt+\pd{L}{v^i}dq^i\right)
=p^*\Theta_L.
\]
and
\[
E_{\hat{L}}=(\Delta w^0)p^*L-w^0p^*L=w^0p^*L-w^0\Delta(p^*L)-w^0p^*L=0.
\]
The other properties follow easily form the first one by using 
$p\circ i=\operatorname{id}$.
\qed

When the Lagrangian $L$ is regular, 
it follows from the preceding proposition that the kernel of 
$\omega_{\hat{L}}$ is 2-dimensional and that $\Delta$ is in the kernel. 
Thus $\hat{L}$ is a singular Lagrangian, but it is easy to see that 
its dynamical equation has solutions defined everywhere and furthermore, 
among them, there are {\sc sode} solutions, 
since $L$ is a type II-Lagrangian
(see~\cite{CCCI-86} for the details).

Locally, if 
$\ds \Gamma=\pd{}{t}+v^i\pd{}{q^i}+f^i(t,q^j,v^j)\pd{}{v^i}$ 
is the solution of the dynamics in the time-dependent formalism, 
then the solution of the dynamics in the homogeneous extended formalism is 
\[
\hat{\Gamma}=w^0\pd{}{x^0}+w^i\pd{}{x^i}+(w^0)^2 f(x^0,x^j,w^j/w^0)\pd{}{w^i}
+\lambda \Delta,
\]
for $\lambda$ an arbitrary function on $\widehat{TE}$.

Once we have transformed our time-dependent problem into an autonomous one, 
and taking into account that the Lagrangian is singular but 
does not generate constraints, we can apply the theory that 
we have developed in the previous sections. We look for a vector field 
$Y$ on $E$ such that its integral curves are also integral curves of 
the dynamical vector fields $\hat{\Gamma}$. 
Since we are interested in integral curves parametrized by time, 
we must chose such vector field $Y$ in the image of the map $i$,
that is, we will take a jet field $X\colon E\to J^1\pi$ 
and the vector field $Y=i\circ X$. Then we have that 
\[
Y^*\theta_{\hat{L}}=Y^*p^*\Theta_L=X^*i^*p^*\Theta_L=X^*\Theta_L,
\]
so that the Hamilton--Jacobi equation amounts to $d(X^*\Theta_L)=0$. 
Locally, the form $X^*\Theta_L$ will be exact, $X^*\Theta_L=dS$, i.e.
\[
X^*\left(-E_Ldt+\pd{L}{v^i}dq^i\right)=\pd{S}{t}dt+\pd{S}{q^i}dq^i.
\]
Thus the Hamilton--Jacobi equation reads in coordinates
\begin{align*}
\pd{S}{t}&=-E_L(t,q^i,X^i)\\
\pd{S}{q^i}&=\pd{L}{v^i}(t,q^i,X^i),\\
\end{align*}
which are the expected expressions of the
Hamilton--Jacobi theory for time-dependent Lagrangian systems.

\subsection{The Hamiltonian formalism}

In time-dependent non-relativistic Hamiltonian Mechanics,
the Hamiltonian is not a function but a section $h$ of a certain bundle.
Given a bundle $\map{\pi}{E}{\Real}$ we consider the affine-dual bundle
$\operatorname{Aff}(J^1\pi,\Real)$, which is canonically isomorphic to $T^*E$,
and the vector bundle $\map{\nu}{J^{1*}\pi\equiv\operatorname{Ver}(\pi)^*}{E}$
dual to the vertical bundle. We have an affine bundle fibration
$\map{\mu}{T^*E}{J^{1*}\pi}$ and a Hamiltonian is a section $h$ of the projection $\mu$. 

Given a Hamiltonian section $\map{h}{J^{1*}\pi}{T^*E}$, the pullback by $h$
of the canonical symplectic form $\omega=-d\theta$ on $T^*E$ defines a 2-form
$\Omega_h=h^*\omega$ on $J^{1*}\pi$. The associated Hamiltonian vector fields
are the solutions $\Gamma_h$ to the equations
\begin{equation}
i(\Gamma_h)\Omega_{h}=0
\qquand
i(\Gamma_h) dt=1.
\end{equation}
It is clear that $\Omega_h=-d\Theta_h$ where $\Theta_h=h^*\theta$.

The relation with the Lagrangian formalism is as follows (see \cite{CaCrIb} for details).
 From the Lagrangian $L$ we can define two maps,
usually called the Legendre transformation $\map{\FL}{J^1\pi}{J^{1*}\pi}$
and the extended Legendre transformations $\map{\hFL}{J^1\pi}{T^*E}$,
related by $\mu\circ\hFL=\FL$. When the Lagrangian is hyper-regular we have that
$\FL$ is invertible and a unique section $h$ of $\mu$ is determined by the equation
$\hFL=h\circ\FL$. When $L$ is regular we must restrict the study to the image of $\FL$.
For simplicity, we will assume that the Lagrangian $L$ is hyper-regular.

Let us consider the homogeneous Lagrangian $\hat{L}\in\cinfty{\widehat{TE}}$
and the Legendre transformation $\map{\FhL}{\widehat{TE}}{T^*E}$ defined by $\hat{L}$.
Then the relation between the Legendre transformation $\FhL$ and $\hFL$
is given by $\hFL=i\circ\FhL$. In coordinates $(x^0,x^i,w^0,w^i)$ in $\widehat{TE}$
and $(x^0,x^i,u,p_i)$ on $T^*E$, the expression of the Legendre transformation is 
\[
\FhL(x^0,x^i,w^0,w^i) =\left(x^0,x^i,-p^*E_L,p^*\left(\pd{L}{v^i}\right)\right),
\]
the composition $\map{\hFL=\FhL\circ i}{J^1\pi}{T^*E}$ is given by 
\[
\hFL(t,x^i,v^i)=\left(t,x^i,-E_L,\pd{L}{v^i}\right),
\]
and composing with the projection $\map{\mu}{T^*E}{J^{1*}\pi}$
we get the map $\map{\FL}{J^1\pi}{J^{1*}\pi}$, which in coordinates reads
\[
\FL(t,x^i,v^i)=\left(t,x^i,\pd{L}{v^i}\right).
\]
Since we are assuming that the Lagrangian $L$ is hyper-regular,
it follows that the Lagrangian $\hat{L}$ is almost-regular,
and we can construct the Hamiltonian formulation.
The kernel of $T\FhL$ is spanned by the Liouville vector field $\Delta$ on $\widehat{TE}$,
and moreover we have $\FhL(\lambda w)=\FhL(w)$ for every $\lambda\neq0$,
so that the image of $\FhL$ coincides with the image of $\hFL$.
Since $\FL$ is invertible, we can identify the image of $\FhL$ with $J^{1*}\pi$,
or better, with the image of $J^{1*}\pi$ by a unique section
$\map{h}{J^{1*}\pi}{T^*E}$ of $\mu$ given explicitly by $h=\hFL\circ\FL^{-1}$. 

Thus, with the same notation as in the general case, we have that
$P=J^{1*}\pi$ and $j_0=h$. If we denote  by $\omega=-d\theta$
the canonical symplectic form on $T^*E$, then the 2-form $\omega_0=j_0^*\omega$
is $\Omega_h=h^*\omega$, that is the differential of $\Theta_h=h^*\theta$.
Following our general theory  for unconstrained singular systems,
we must look for a section $\alpha$ such that
$\alpha^*\Theta_h$ is locally an exact form $dS$.
In coordinates $h(t,x^i,p_i)=(t,x^i,-H(t,x^i,p_i),p_i)$ and hence 
\[
\alpha^*\Theta_h=\alpha_i dq^i-(H\circ\alpha)\, dt =\pd{S}{t}dt+\pd{S}{x^i}dx^i,
\]
from where we get 
\[
\alpha_i=\pd{S}{x^i} 
\qquand 
\pd{S}{t}+H(t,x^i,\alpha_i)=0,
\]
or equivalently
\[
 \pd{S}{t}+H\left(t,x^i,\pd{S}{x^i}\right)=0,
\]
which is the classical time dependent Hamilton--Jacobi equation.

\section{Distance on a Riemann manifold: the free relativistic particle}
\protect\label{frp}

\subsection{General features}

We consider a Riemannian or semi-Riemannian manifold $(Q,g)$
and the Lagrangian $L(v)=\sqrt{g(v,v)}$. In the semi-Riemannian case we restrict $v$
to be time-like,
i.e., $g(v,v)>0$. In particular, if $g$ is the Lorentz metric,
this Lagrangian models a free relativistic particle on the manifold~$Q$.

\paragraph{Lagrangian dynamics}

The Lagrangian $L$ is singular. In fact, it is homogeneous 
of degree one, hence the energy function vanishes identically $E_L=0$.
The Cartan 1-form is given by
\[
\theta_L(U)=\frac{g(v,w)}{\sqrt{g(v,v)}}
\]
for all $U\in T(TQ)$, where $v=\tau_{TQ}(U)$ and $w=T_v\tau_Q(U)$. 
The kernel of the Cartan 2-form $\omega_L$ is generated by the geodesic spray 
$\Gamma$ and the Liouville vector field $\Delta$. 
There exists underdetermined global second-order dynamics given by 
$\Gamma+\lambda\Delta$, for any function $\lambda\in\cinfty{TQ}$. See~\cite{Ca-90,Gr00}.

\paragraph{Hamilton--Jacobi equation}

Let $X$ be a nowhere vanishing  vector field on $Q$,
and everywhere time-like in the semi-Riemannian case.
 From the expression of $\theta_L$ above we immediately have that 
$$
X^*\theta_L=\frac{1}{\sqrt{g(X,X)}}\,\gX,
$$
where we have denoted by $\gX$ the 1-form on $Q$ such that $\pai{\gX}{Y}=g(X,Y)$
for all vector fields $Y$ on $Q$. If we define $\UX$ as the unitary vector field
in the direction of $X$, that is 
\[
\UX=\frac{1}{\sqrt{g(X,X)}}X,
\]
then we have that $X^*\theta_L=\gUX$.

Since the energy function vanishes identically,
the Hamilton--Jacobi equation reduces to $d(X^*\theta_L)=0$.
Let us find an alternative expression for this condition in terms of the
Levi-Civita connection associated with the metric.

\begin{proposition}
A time-like vector field $X\in\vf{Q}$ is a solution of the Hamilton--Jacobi equation
of the Lagrangian $L(v)=\sqrt{g(v,v)}$ if, and only if,
\begin{enumerate}
\item the distribution $X^\perp$ is integrable, and 
\item $\nabla_XX=\lambda X$ for some function $\lambda\in\cinfty{Q}$.
\end{enumerate}
\end{proposition}
\begin{proof}
If $X$ is a solution of the Hamilton--Jacobi equation, i.e. $d(X^*\theta_L)=0$,
we have that $d\gUX=0$, hence $X^\perp=(\gUX)^\circ$ is an integrable distribution.

Observe that, for every vector field $Z\in\vf{Q}$,
using the Levi--Civita connection associated with the metric $g$,
since this connection is torsion-free,
the exterior differential can be calculated by skew-symmetrization 
of the covariant differential, and thus
\begin{equation}
d\gX(Y,Z)=\nabla_Y\gX(Z)-\nabla_Z\gX(Y)=g(\nabla_YX,Z)-g(\nabla_ZX,Y)\, .
\label{formulilla}
\end{equation}
Then we have
\begin{align*}
0=(d\gUX)(\UX,Z)
&=g(\nabla_\UX \UX,Z)-g(\nabla_Z\UX,\UX)\\
&=g(\nabla_\UX \UX,Z)-\frac{1}{2}\nabla_Z(g(\UX,\UX))
=g(\nabla_\UX \UX,Z)
\end{align*}
then $\nabla_\UX\UX=0$, and hence $\nabla_XX=\lambda X$
with $\lambda=\nabla_X(\ln\sqrt{g(X,X)})$. This proves the direct statement.

Conversely, assume that $X$ satisfies conditions $(1)$ and $(2)$.
We will prove that $d\gUX=0$, so that $d(X^*\theta_L)=0$.
Taking the derivative $\nabla_X$ of $g(X,X)$ we find that the function $\lambda$
is given by the relation $\nabla_X\sqrt{g(X,X)}=\lambda\sqrt{g(X,X)}$,
from where $\nabla_\UX\UX=0$ follows.

On the other hand, if the distribution $X^\perp$ is integrable, there
exists locally a nowhere vanishing function $\varphi$ such that $d(\varphi\gUX)=0$.
First we will prove that $d\varphi=(\nabla_\UX\varphi)\gUX$. Indeed, for every $Z\in\vf{Q}$,
\begin{align*}
0=d(\varphi\gUX)(\UX,Z)
&=\nabla_\UX(\varphi\, g(\UX,Z))-\nabla_Z(\varphi\, g(\UX,\UX))-\varphi\, g(\UX,[\UX,Z])\\
&=(\nabla_\UX\varphi)g(\UX,Z)-\nabla_Z\varphi
\end{align*}
where we have used that $g(\UX,\UX)=1$ and 
\[
g(\UX,[\UX,Z])
=g(\UX,\nabla_\UX Z)-g(\UX,\nabla_Z\UX)
=g(\UX,\nabla_\UX Z)-\frac{1}{2}\nabla_Z(g(\UX,\UX))
=g(\UX,\nabla_\UX Z)
\]
Therefore $\nabla_Z\varphi=(\nabla_\UX\varphi)g(\UX,Z)$, for every $Z\in\vf{Q}$,
which proves that $d\varphi=(\nabla_\UX\varphi)\gUX$. But we have
\[
0=d(\varphi\gUX)=d\varphi\wedge\gUX+\varphi\,d\gUX=\varphi\,d\gUX,
\]
that is $d\gUX=0$, and hence $d(X^*\theta_L)=0$.
\qed
\end{proof}

\subsection{Alternative Lagrangian description}

It is well known~\cite{Bergman} that $L(v)=\frac{1}{2}g(v,v)$ 
provides a Lagrangian description of a free motion in
a Riemannian manifold, that is, the geodetic spray $\Gamma$, which is, moreover, regular.
It is interesting to compare the solutions of the Hamilton--Jacobi equation
for this Lagrangian with the above one.

The Lagrangian $L$ is homogeneous of degree two. 
Therefore, we have that $E_L=L=\frac{1}{2}g(v,v)$. 
The Cartan 1-form is given by $\theta_L(W)=g(v,w)$ for $W\in T_v(TQ)$,
where $w=T_v\tau_Q(W)$, so that $X^*\theta_L=\gX$. 
Then the Hamilton--Jacobi equation is
$$
d\gX=0\qquand g(X,X)=c,
$$
for some constant $c\geq0$. 
For $c=0$ we have the trivial solution $X=0$, and for $c>0$ 
we can rescale $X$ to $X/\sqrt{c}$,
so that we can consider only the case $c=1$, 
which means that we can restrict our study to the case that $X$ is a unit vector field.

\begin{proposition}
A unit vector field $X$ is a solution of the Hamilton--Jacobi equation
for the Lagrangian $L(v)=\frac{1}{2}g(v,v)$
if, and only if, $X^\perp$ is integrable and $\nabla_XX=0$.
\end{proposition}
\proof
By (\ref{formulilla}), we have
$d\gX(Y,Z)=g(\nabla_YX,Z)-g(\nabla_ZX,Y)$, 
so that the Hamilton--Jacobi equation is equivalent to
$$
g(\nabla_YX,Z)=g(\nabla_ZX,Y).
$$
If we take $Y=Z=X$ the condition is identically satisfied. 
If we take $Y=X$ and $Z\in X^\perp$, we have $g(\nabla_XX,Z)=0$, 
thus the vector field $X$ satisfies that $\nabla_XX=\lambda X$ 
for some function $\lambda\in\cinfty{Q}$. 
But from the normalization condition $g(X,X)=1$ we have that 
$\lambda=g(\nabla_XX,X)=\frac{1}{2}\nabla_X[g(X,X)]=0$, 
so that $\nabla_XX=0$. Finally, for $Y,Z\in X^\perp$ we have
$g(\nabla_YX,Z)=g(\nabla_ZX,Y)$, which, as above, is equivalent to 
$g(X,[Y,Z])=0$.
\qed

\paragraph{Remarks}
\begin{enumerate}
\item
The condition $\nabla_XX=0$ gives the generalized solution of the Hamilton--Jacobi problem,
and together with the integrability of $X^\perp$
give the classical Hamilton--Jacobi solution.
\item
Recall that a vector field $X$ satisfying that $\nabla_XX=0$
is called a {\sl geodetic vector field},
and its integral curves are geodesics parametrized by arc length.
If we reparametrize the curves we have the vector field $\bar{X}=fX$ 
for some function $f$ nowhere vanishing. Therefore 
$\nabla_{\bar{X}}\bar{X}=f(Xf)X$, so that 
$\nabla_{\bar{X}}\bar{X}=\lambda\bar{X}$ with $\lambda=f(Xf)$.
Notice the relation between the unit length parametrization 
in the regular case with the projective theory in the singular case. 
In the regular case, the vector field $X$ must be unitary 
in order to have integral curves parametrized by arc-length.

\item 
The interpretation of the above results is (in both cases) as follows:
the vector $X$ points in the direction of propagation of the rays,
and the orthogonal distribution to $X$ is the tangent to the wavefront.
Wavefronts are manifolds, so the orthogonal distribution to $X$ is integrable. 
Furthermore the rays are the geodesics of the metric, 
and therefore the vector field $X$ must be a geodetic vector field.
\end{enumerate}

\section{Free motion on Lie groups, rigid bodies and the electron monopole system}
\protect\label{ems}

In this section we wish to show that the notion of generalized solution
is the only one available in generic situations, because solutions
in terms of characteristic functions are not available globally either
for topological reasons or because of invariance requirements.
We are going to present a simplified approach to dynamics on
Lie groups (see~\cite{CLM}), since we wish to isolate
the main conceptual aspects of the Hamilton--Jacobi problem on spaces
with nontrivial topology, although parallelizable.

\subsection{Free motion on Lie groups}

By free motions on a Lie group $G$ we mean motions
associated with equations of motion analogous to the equation $d^2x/dt^2=0$,
which are written in some
affine space. Thus for simplicity we consider our group realized as a group
of matrices $g\in GL(n,\R)$. The equations of free motion will be written as 
$$\frac d{dt}\bigl(g^{-1}(t)\,\dot g(t)\bigr)=0\,.
$$
These differential equations admit a Lagrangian description
in terms of a Lagrangian function  
$$
L(g,\dot{g})=\frac 12 {\rm Tr\,}\left[(g^{-1}\,\dot g)^2\right]\,.
$$
The geometrical objects associated with $L$ are simply written
$$
\theta_L(g,\dot{g})= {\rm Tr\,}\left[(g^{-1}\,\dot g)\,(g^{-1}\,d g)\right]\,,\quad
  \omega_L=-d\theta_L,\quad E_L=L\,.
$$

We will show that every left-invariant vector field $X$ provides us
with a solution of the generalized Hamilton--Jacobi problem.
So, let us consider $X\in\vf_L{G}$. Denote by $\xi$
the value of $X$ at the identity $e$ in the group $G$, that is $\xi=X(e)\in\g$.
In this way, we have that $X(g)=g\xi$, or $g^{-1}X(g)=\xi$.

On the one hand, it is clear that the pullback of the energy is constant:
\[
(X^*E_L)(g)=(X^*L)(g)=\frac{1}{2}\Tr[(g^{-1}X(g))^2]=\frac{1}{2}\Tr(\xi^2).
\]
On the other hand, the pullback of the symplectic form does not vanish.
Indeed, we calculate $\pai{X^*\theta_L}{Y}$ for a vector field
$Y\in\vf{G}$, which we may take to be left-invariant, $Y(g)=g\zeta$,
for some $\zeta\in\g$:
\[
\pai{X^*\theta_L}{Y}(g)
=\frac{d}{dt}L(X(g)+tY(g))\at{t=0}
=\frac{1}{2}\frac{d}{dt}\Tr[(\xi+t\zeta)^2]\at{t=0}
=\Tr(\xi\zeta).
\]
Thus the differential evaluated on two left-invariant vector fields $Y_1, Y_2$,
 $Y_1(g)=g\zeta_1$ and $Y_2(g)=g\zeta_2$, is
\begin{align*}
d(X^*\theta_L)(g)(Y_1,Y_2)
&=Y_1(g)(\Tr(\xi\zeta_2))-Y_2(g)(\Tr(\xi\zeta_1))- \Tr(\xi\,g^{-1}[Y_1,Y_2](g)))\\
&=0-0-\Tr(\xi\,[\zeta_1,\zeta_2]),
\end{align*}
so that $(X^*\omega_L)(Y_1,Y_2)=\Tr(\xi\,[\zeta_1,\zeta_2])$. 

It follows that $X$ is not a solution of the Hamilton--Jacobi problem
in the standard sense (except for Abelian groups).
Nevertheless $X$ is a solution of the generalized Hamilton--Jacobi problem,
since $i_X(X^*\omega_L)=0$:
\[
i_X(X^*\omega_L)(Y)
=\Tr(\xi\,[\xi,\zeta])
=\Tr(\xi^2\zeta-\xi\zeta\xi)=0,
\]
where use has been made of the properties of the trace.

Therefore, the left trivialization provides a diffeomorphism
$\map{\Phi}{G\times\g}{TG}$, given by $\Phi(g,\xi)=T_eL_g(\xi)$,
such that for every $\xi\in\g$ we get a solution of the generalized problem.
In other words we have a complete solution of the generalized Hamilton--Jacobi problem,
where the parameter space $\Lambda$ is the Lie algebra, $\Lambda=\g$.

The associated constant of the motion
$\map{F=\mathrm{pr}_2\circ\Phi^{-1}}{TG}{\g}$ is explicitly given by
$F(g,\dot{g})=g^{-1}\dot{g}$, or in other words $F(g,g\xi)=\xi$.
We can identify $\g$ with $\g^*$ via the trace operation,
that is we identify $\nu\in\g^*$ with $\xi\in\g$ if
$\pai{\nu}{\zeta}=\Tr(\xi\zeta)$ for every $\zeta\in\g$.
Under this identification, we get a map $\map{\mu}{TG}{\g^*}$ given by 
\[
\pai{\mu(g,\dot{g})}{\zeta}=\Tr(g^{-1}\dot{g}\zeta),
\]
which is the momentum map for the left action of $G$ on $TG$.

Notice that we have exploited the left-invariance of the Lagrangian function.
Furthermore $L$ is also right invariant, since we can write it in the form
$L(g,\dot{g})=\frac{1}{2}\Tr[(\dot{g}g^{-1})^2]$.
Therefore we can also define a second foliation by taking the right-invariant vector fields.

Finally, it should be remarked that whenever $G$ is a compact Lie group we cannot have
functions on $G$ whose differentials are never vanishing, therefore any
invariant foliation (solutions of the generalized Hamilton--Jacobi problem)
could never be associated with some $dW$, for some function $W:G\times
\Lambda\to \R$.

\subsection{Rigid bodies}

Consider a (generalized) rigid body defined on a configuration Lie group $G$
with symmetric inertia tensor $\map{I}{\g}{\g^*}$.
We will analyze it in the Hamiltonian formalism on the cotangent bundle $T^*G$.
The Hamiltonian function $H\in\cinfty{T^*G}$ is 
\[
H(\lambda_g)=\frac{1}{2}\pai{T^*_eL_g(\lambda_g)}{I^{-1}T^*_eL_g(\lambda_g)},
\]
where $\pai{\ }{\ }$ is the standard pairing.
We will show that every right-invariant 1-form $\alpha\in\Omega^1(G)$
is a solution of the generalized Hamilton--Jacobi problem.
If $\mu\in\g^*$ is the value of $\alpha$ at the identity $e\in G$,
then we have that $\alpha(g)=T^*_gR_{g^{-1}}(\mu)$.
The value of $\alpha$ on a right-invariant vector field $Y=T_e R_g(\zeta)$ is constant,
\[
\pai{\alpha}{Y}(g)=\pai{T^*_gR_{g^{-1}}\mu}{T_eR_g(\zeta)}=\pai{\mu}{\zeta},
\]
and hence the differential of $\alpha$ over two right-invariant vector fields
$Y_1$ and $Y_2$ is
\begin{align*}
d\alpha(Y_1,Y_2)(g)
&=Y_1(g)\pai{\alpha}{Y_2}-Y_2(g)\pai{\alpha}{Y_1}-\pai{\alpha}{[Y_1,Y_2]}(g)\\
&=Y_1(g)\pai{\mu}{\zeta_2}-Y_2(g)\pai{\mu}{\zeta_1}-\pai{\mu}{-[\zeta_1,\zeta_2]}\\
&=0-0+\pai{\mu}{[\zeta_1,\zeta_2]}.
\end{align*}
Therefore, for every $\zeta_1,\zeta_2\in\g$ we have 
\[
d\alpha(g)(T_eR_g(\zeta_1),T_eR_g(Z\zeta_2))
=\pai{\mu}{[\zeta_1,\zeta_2]}.
\]

The fiber derivative of the Hamiltonian is given by 
$\FH(\lambda_g)=T_eL_g\,I^{-1}\,T^*_eL_g(\lambda_g)$,
for every $\lambda-g\in T_g^*G$. Indeed
\[
\pai{\lambda'_g}{FH(\lambda_g)}
=\frac{d}{dt}H(\lambda_g+t\lambda'_g)\at{t=0}
=\pai{T^*_eL_g(\lambda'_g)}{I^{-1}T^*_eL_g(\lambda_g)}
=\pai{\lambda'_g}{T_eL_gI^{-1}T^*_eL_g(\lambda_g)}.
\]
Therefore, the vector field $X\in\vf{G}$ associated to $\alpha$ is
\[
X(g)
=\FH(\alpha(g))
=\FH(T^*_gR_{g^{-1}}(\mu))
=T_eL_gI^{-1}(Ad^*_g(\mu)),
\]
The contraction of $d\alpha$ with $X$ is given by 
\begin{align*}
(i_X d\alpha)(g)(TR_g(\zeta))
&=d\alpha(g)(X(g),TR_g(\zeta))\\
&=d\alpha(g)(T_eR_g(T_gR_{g^{-1}}X(g)),TR_g(\zeta))\\
&=\pai{\mu}{[Ad_gI^{-1}Ad_g^*\mu,\zeta]},
\end{align*}
where we have used that
\[
T^*_gR_{g^{-1}}(X(g))
=T^*_gR_{g^{-1}}T_eL_gI^{-1}Ad^*_g\mu
=Ad_gI^{-1}Ad^*_g\mu.
\]

Furthermore, the pullback of the Hamiltonian by $\alpha$ is 
\[
(\alpha^*H)(g)=\frac{1}{2}=
\pai{T^*_eL_g(T^*_gR_{g^{-1}}(\mu))}{I^{-1}T^*_eL_g(T^*_gR_{g^{-1}}(\mu))}
=\frac{1}{2}\pai{Ad^*_g(\mu)}{I^{-1}Ad^*_g(\mu)}.
\]
To calculate its differential evaluated at $T_eR_g(\zeta)$,
we consider its integral curve $\gamma(t)=\exp(t\zeta)g$ through the point $g$ and hence
\begin{align*}
\pai{d(\alpha^*H)(g)}{T_eR_g(\zeta)}
&=\frac{d}{dt}(\alpha^*H)(\gamma(t))\at{t=0}\\
&=\pai{\frac{d}{dt}Ad^*_{\exp(t\zeta)g}\mu\at{t=0}}{I^{-1}Ad^*_g\mu}\\
&=\pai{Ad^*_gad^*_\zeta\mu}{I^{-1}Ad^*_g\mu}\\
&=\pai{ad^*_\zeta\mu}{Ad_gI^{-1}Ad^*_g\mu}\\
&=\pai{\mu}{ad_\zeta  Ad_gI^{-1}Ad^*_g\mu}
\end{align*}
and finally adding both terms we get
\[
\bigl(i(X)d\alpha+d(\alpha^*H)\bigr)(T_eR_g(\zeta))
=\pai{\mu}{[Ad_gI^{-1}Ad_g^*\mu,\zeta]}+\pai{\mu}{ad_\zeta Ad_gI^{-1}Ad^*_g\mu}
=0.
\]

Thus we have a complete solution of the generalized Hamilton--Jacobi problem,
$\map{\Phi}{G\times\g^*}{T^*G}$ explicitly given by the inverse
of the right trivialization map, $\Phi(g,\mu)=T^*_gR_{g^{-1}}(\mu)$.
The associated constant of the motion is $\map{F=\mathrm{pr}_2\circ\Phi^{-1}}{T^*G}{\g^*}$,
which is the momentum map $F=J_L$ associated to the left action of $G$ on $T^*G$, that is, 
\[
F(\lambda_g)=T^*_eR_g(\lambda_g).
\]

As the theory predicts, if $g(t)$ is an integral curve of the vector field $X$,
i.e. it satisfies $\dot{g}(t)=T_eL_{g(t)}I^{-1}(Ad^*_{g(t)}\mu)$,
then $\Omega=g^{-1}\dot{g}$ is given by $I\Omega=Ad^*_g\mu$ (with $\mu\in\g^*$ constant)
and hence it satisfies the differential equation
$I\dot{\Omega}=-ad^*_\Omega Ad^*_g\mu=-ad^*_\Omega(I\Omega)$,
which is the Euler equation.

\subsection{The electron monopole system}

The equations of motion for  a charged particle with electric charge $e$
moving in the external magnetic field of  a monopole with magnetic charge $g$ 
 are described by the following second order vector field in $Q = \R^3 - \{0\}$ 
$$\Gamma=v^j\, \pd{}{q^j}+\frac n{r^3}  \, \epsilon_{ijk}\, x^j\, v^k\,
\pd{}{v^i}\ ,
$$
 where  $\epsilon_{ijk}$ is the completely skew-symmetric Levi Civita tensor,
i.e. such that $\epsilon_{123}=1$ and with
$$ n = \frac{e\,g}{4\,\pi\, m}\,.$$

The vector field $\Gamma$ admits a symplectic description with the symplectic
structure (see e.g \cite{MR88})
$$\omega=dx^i\land dv^i-\frac n{2\,r^3}\,\epsilon_{ijk}\,x^i\,dx^j\land dx^k\ ,
$$
and Hamiltonian function 
$$H=\frac 12 v^j\, v^j\,.
$$
Because $\omega$ is closed but not exact, $\omega$ cannot be written as a
Lagrangian 2-form $\omega_L$. It is however possible to write it as a 
Lagrangian 2-form locally by using a local Lagrangian.

In addition to the Hamiltonian function, the dynamical system admits other
constants of the motion associated with the rotational symmetry group; they are
$$l_i=\epsilon_{ijk}\,x^j\, v^k+\frac {n\, x^i}r\,.
$$
They are made up of the expected components of the orbital angular momentum
plus the ``helicity term'' $n\, x^j/r$.

It is possible to find local solutions of the standard Hamilton--Jacobi
equation by using constants of the motion $H$, ${\bf l}^2$ and $l_3$, for
instance. We may solve for the velocities, and by replacing them in $\Gamma$ we
find a 3-parameter family of vector fields defined on some open submanifold of
$ \R^3 - \{0\}$.

It should be noticed, however, that it is not possible to find globally defined
vector valued solutions, because if we denote the
sought solution by $Y=Y^j\, \pd{}{x^j}$, we would have 
$$
Y^*(dx^i\wedge dv^i)=\frac n{2\, r^3}\epsilon_{jki}\,x^j\,dx^k\wedge dx^i\,,
$$
which is not possible because the left hand side is exact while the right hand
side is not.

Nevertheless, it is possible to describe the electron monopole system as a
reduction of a globally defined Lagrangian system with a singular Lagrangian 
but without secondary constraints. To this end we replace the
configuration space $Q =  \mathbb{R}^3-\{0\} \approx S^2\times \mathbb{R}_+$
with a covering by replacing $S^2$ with $S^3$ in the product of manifolds.
The new configuration space will be $SU(2,\mathbb{C})\times \mathbb{R}_+$.

The covering map $\pi:SU(2,\mathbb{C})\to S^2$ is given by the following
construction. Let $(x^1,x^2,x^3)$ be the coordinates in $\mathbb{R}^3-\{0\}$
and let $\widehat x^j=x^j/r\in S^2$, so that they satisfy
$\widehat x^j\,\widehat x^j=1$. Now we describe $\mathbb{R}^3$ in terms of the
$2\times 2$ traceless Hermitian matrices using as a basis Pauli matrices, we
have 
$$M=\vec x\cdot \vec \sigma=
\begin{bmatrix}
x^3&x^1-ix^2\\x^1+ix^2&-x^3
\end{bmatrix}.
$$

Now we describe our covering map by introducing the following matrices to
describe $ \mathbb{R}^4$
$$
{\bf s}=
y^0\,I+i\vec y\cdot \vec \sigma=
\begin{bmatrix}
y^0+iy^3&y^2+iy^1
\\-y^2+iy^1&y^0-iy^3
\end{bmatrix},
$$
and setting $\pi:\mathbb{R}^4\to \mathbb{R}^3$ by means of
$$\pi ({\bf s})={\bf s}\,\sigma_3\, {\bf s}^{\dag}=\vec x\cdot \vec \sigma\,.
$$
This map is also known as the Kustaanheimo-Stiefel map
(for a classical and quantum version of this map
see the recent papers~\cite{Avanzo1, Avanzo2}).

This relation makes sense because both sides are traceless
Hermitian matrices and Pauli matrices are a basis for the real linear space of
Hermitian matrices with zero trace. We notice that ${\bf s}$ represent
elements of $SU(2,\mathbb{C})$ when the constraint 
$${\bf s}\,{\bf s}^{\dag}=((y^0)^2+(y^1)^2+(y^2)^2+(y^3)^2)\, I=\det s\, I=I$$
is imposed.

To spell out the way  $(x^1,x^2,x^3)$ depends on $(y^0,y^1,y^2,y^3)$, i.e. 
the pull-back of coordinate functions from $\mathbb{R}^3-\{0\}$  to
$ \mathbb{R}^4-\{0\}$, we notice that 
$$
{\bf s}^{\dag}=\begin{bmatrix} y^0-iy^3&-y^2-iy^1\\y^2-iy^1&y^0+iy^3
\end{bmatrix}
$$
so that 
$${\bf s}\,\sigma_3\,{\bf s}^{\dag}=\begin{bmatrix}
y^0+iy^3&y^2+iy^1
\\-y^2+iy^1&y^0-iy^3
\end{bmatrix}
\begin{bmatrix}1&0\\0&-1\end{bmatrix} 
\begin{bmatrix} y^0-iy^3&-y^2-iy^1\\y^2-iy^1&y^0+iy^3
\end{bmatrix}
$$
which is given by 
$$
{\bf s}\,\sigma_3\,{\bf s}^{\dag}=\begin{bmatrix}
(y^0)^2+(y^3)^2-(y^1)^2-(y^2)^2&2(y^0\,y^1+y^2\,y^3)-2i(y^0\,y^2-y^1\,y^3)\\
2(y^0\,y^1+y^2\,y^3)+2i(y^0\,y^2-y^1\,y^3)&-(y^0)^2-(y^3)^2+(y^1)^2+(y^2)^2
\end{bmatrix}
$$
provides us with 
\begin{eqnarray}
x^1&=& 2(y^0\,y^1+y^2\,y^3)\cr
x^2&=& 2(y^0\,y^2-y^1\,y^3)\cr
x^3&=& (y^0)^2+(y^3)^2-(y^1)^2-(y^2)^2.\nonumber
\end{eqnarray}

Now we find the tangent map of the covering map 
$$T\pi:T(SU(2,\mathbb{C})\times \mathbb{R}_+)\to T(S^2\times \mathbb{R}_+)\,,$$
more explicitly
\begin{eqnarray}
v^1&=&2(y^0u^1+u^0 y^1+u^2y^3+y^2 u^3)\cr
v^2&=&2(y^0u^2+u^0 y^2-u^3y^1+y^3 u^1)\cr
v^3&=&2(y^0u^0+ u^3y^3-y^1u^1-y^2 u^2)\nonumber
\end{eqnarray}
and the pull-back of the 2-form 
$$\omega=dx^i\land dv^i-\frac n{2\,r^3}\,\epsilon_{ijk}\,x^i\,dx^j\land dx^k\ ,
$$
namely, $(T\pi)^*\omega$, will be exact and moreover Lagrangian 
$$
(T\pi)^*\omega=\omega_L,
$$
with Lagrangian $L$ on $T(SU(2,\mathbb{C})\times \mathbb{R}_+)$ given by 
$$L=\frac 12 T\pi^*(v^jv^j)+i\, n\, {\rm Tr\,}\sigma_3\, {\bf s}^{-1}\, \dot {\bf s}\,.
$$

The fibering map $\pi:SU(2,\mathbb{C})\times \mathbb{R}_+\to S^2\times
\mathbb{R}_+$ is actually a principal bundle projection with group $U(1)$ given
by 
$$U(1)=\{\exp(i\, t\, \sigma_3)\mid t\in \mathbb{R}\}\,,
$$
and acting on $SU(2,\mathbb{C})$ on the right. The tangent bundle group
$TU(1)=U(1)\otimes \mathbb{R}$ will now be the structure group of the tangent
bundle 
$T(SU(2,\mathbb{C})\times \mathbb{R}_+)\to T(S^2\times
\mathbb{R}_+)$.

Within the notation we have used, the left-invariant vector field along 
$\sigma_3$
generator of the $U(1)$ action  is 
$$X^3=y^0\pd {}{y^3}-y^3\pd {}{y^0}-\epsilon _{3ij}\, y^i\pd {}{y^j}.
$$
while the infinitesimal generator of $TU(1)$ will be the tangent lift
$$
(X^3)^T=X^3+\dot y^0\pd{}{\dot y^3}-\dot y^3\pd{}{\dot y^0}-\epsilon _{3ij}\,
\dot y^i\pd {}{\dot y^j}
$$
and the vertical lift 
$$
(X^3)^v=y^0\pd{}{\dot y^3}- y^3\pd{}{\dot y^0}-\epsilon _{3ij}\,
 y^i\pd {}{\dot y^j}\,.
$$

By using the pull-back of constants of the motion from
$T(\mathbb{R}^3-\{0\})$ to $T( \mathbb{R}^4-\{0\})$,
we will find generalized extended space. On the
8-dimensional carrier space they will define submanifolds of codimension
three. If the constants of the motion used are pairwise in involution, the
invariant submanifold will be isotropic, otherwise we will give rise to
solutions of the generalized Hamilton--Jacobi problem.

\section{Conclusions and outlook}

In this paper we show that to deal with bi-Hamiltonian systems in the
Hamilton--Jacobi setting it is convenient to introduce generalized solutions,
i.e. invariant submanifolds (or foliations)  with dimension equal to the
dimension of the configuration space without the requirement of
Lagrangianity. Thus the associated {\sc pde} will have solutions given by
vector valued functions. When the ``Lagrangian'' requirement is made, these
functions will be the coefficients of an exact 1-form, and we recover the
standard {\sc pde} for the principal function $W$ or the characteristic function
$S$. The link via the Feynman approach to quantum mechanics between these
solutions and the phase of the wave function seems to suggest that only
invariant foliations with Lagrangian leaves with respect to the admissible
alternative symplectic structure should be accepted.

According to von Neumann's representation theory we would have to accept as
Hilbert spaces the space of square integrable functions defined on some
invariant Lagrangian submanifold (according to the chosen symplectic 
structure).
This raises the problem of the selection of the appropriate ``Lebesgue
measure'' and how to compare the descriptions on these alternative Hilbert
spaces. These aspects will be taken up elsewhere.

Formulation of the Hamilton--Jacobi theory on the tangent bundle
in terms of the Lagrangian formalism, prepares us ready to consider the problem of 
Hamilton--Jacob theory connected with degenerate Lagrangians (gauge theories)
in full generality. Extension of the ideas in this paper for
Lagrangian and Hamiltonian systems on Lie algebroids~\cite{LMLA,LSDLA}
is also worthy of study. These aspects should be addressed in the future.

\subsection*{Acknowledgments}

Support of projects {\sc MEC (Spain)} BFM-2003-02532, 
BFM-2002-03773, BFM-2002-03493, MTM2005-04947, and FPA-2003-02948 and  CO2-399
is acknowledged.
We wish to thank Mr. Jeff Palmer for his
assistance in preparing the English version of the manuscript.



\end{document}